\documentclass[amsmath,amssymb,twocolumn,dvipdfm]{jpsj3} 
\usepackage{bm}
\usepackage{graphicx,color}%

\newcommand{\rd}[1]{#1}
\newcommand{\bl}[1]{#1}

\newcommand{\crd}[1]{#1}
\newcommand{\dark}[1]{#1}

\makeatletter
    
    \@addtoreset{equation}{section}
  \makeatother

\title{Impurity Effects on Superconductivity on Surfaces of Topological Insulators}
\author{  
Yuto \textsc{Ito}, Youhei \textsc{Yamaji}, and Masatoshi \textsc{Imada}
}

\inst{
Department of Applied Physics, University of Tokyo, Bunkyo, Tokyo 113-8656, Japan
}
\date{today}

\abst
{
A \rd{two-dimensional} superconductor (SC) on surfaces of topological insulators (TIs) is a mixture of $s$-wave and
helical $p$-wave components when
induced by $s$-wave interactions, since spin and momentum are correlated.
\crd{On the basis of} the Abrikosov-Gor'kov theory,
we reveal that
unconventional SCs on \crd{the} surfaces of TIs are stable against \rd{time-reversal symmetric (TRS)} impurities within
\crd{a} region of small impurity concentration.
Moreover\crd{,} we analyze the stability of the SC on \crd{the} surfaces of TIs against impurities beyond the perturbation theory
by solving the real-space Bogoliubov-de Gennes equation for an effective tight-binding model of a TI.
We find that the SC is stable against strong TRS \bl{impurities.}
\crd{The b}ehaviors of bound states around an impurity \rd{suggest} that the SC on \crd{the} surfaces of TIs is not a topological SC.
}
\kword{topological insulator, helical Dirac electron, unconventional superconductivity, impurity scattering, time reversal symmetry, spin orbit interaction, Abrikosov-Gor'kov theory, Bogoliubov-de Gennes equation, impurity induced state}
\begin{document}
\maketitle
\section{Introduction}
\rd{Three-dimensional (3D) topological insulators (TIs) have two-dimensional (2D) surface states (SSs) topologically protected by the time reversal symmetry (TRS)\cite{cit:RMP82_3045,cit:RMP83_1057}.
The existence of
SSs \bl{is} characterized by $Z_2$ topological invariants, which are determined
\crd{by the} band structure of the bulk\cite{cit:PRL95_146802,cit:PRL98_106803,cit:PRB75_121306,cit:PRB79_195322}. 
In most cases, strong spin orbit interactions (SOIs) play important roles in constructing topologically nontrivial band structure\crd{s} and \crd{induce} TIs.
By SOIs, helical spin structure\crd{s} in momentum space \crd{are} observed, i.e., helical Dirac electrons are generated.
Helical Dirac electrons in TIs have been verified \bl{through} observations of energy dispersions of Dirac SSs by angle-resolved photoemission spectroscopy (ARPES)\cite{cit:Nature452_970,cit:Nphys5_398,cit:Nature460_1101,cit:PRL105_076802}.
}


Superconductivity on \crd{the} surfaces of TIs attracts attention as one \crd{type} of 2D unconventional superconductivity.
\crd{An} unconventional
\crd{superconductor} (SC) is a mixture of $s$-wave and helical $p$-wave components, since in helical Dirac electron systems, spin and momentum are correlated\cite{cit:PRB81_184502}.
Such a SC is possibly induced by a proximity effect from an $s$-wave SC to a TI\cite{cit:PRL100_096407,cit:PRB81_241310} or, in other words, by an $s$-wave attractive interaction\cite{cit:PRB81_184502}, when the Fermi energy is away from the Dirac point.
This SC formally resembles a spinless chiral $p$-wave SC that breaks TRS\cite{cit:PRB61_10267} in the representation where the Dirac electron dispersion is diagonalized. 
The difference is that the SC on \crd{the} surfaces of TIs does not break TRS.
\crd{From} that similarity of the two SCs,
\crd{an} unconventional SC on the SSs of TI is proposed
\crd{for} application
\crd{to} quantum computations using Majorana bound states caused by the proximity effect between a superconductor and the surface states of TI\cite{cit:PRL100_096407}.
Introducing superconductivity
\crd{into} the surfaces of TIs has been a challenge
\crd{in} experimental research.
For example, toward this goal, a SC has been realized in Cu-doped $\mathrm{Bi_2Se_3}$\cite{cit:PRL104_057001,cit:Nphys6_855},
\crd{al}though \crd{the} existence of
\crd{a} surface SC is not confirmed yet.
Recently, it
\crd{has been} reported that superconductivity is introduced
\crd{into} $\mathrm{Bi_2Se_3}$ thin films by the superconductivity proximity effect\cite{cit:arXiv1112.1772}.

In realizing such a SC, \crd{the} stability of the SC is an important problem. 
In particular, impurity effects are relevant to the stability since surfaces on TI frequently contain disorders such as defects or impurity potentials.
Moreover,
\crd{on} surfaces of TIs, there are facets or steps with disordered boundaries.
For example, by scanning tunneling spectroscopy (STS) studies
\crd{of the} surfaces of $\mathrm{Bi_2Te_3}$, one of TI\crd{s}, an abrupt change
\crd{in} the local density of states (LDOS) was observed near a step structure\cite{cit:PRL104_016401}.

Moreover, studies
\crd{of} the stability against impurities allow us to clarify
\crd{the} fundamental physics of
\crd{a} SC on \crd{the} surfaces of TIs since the stability of \crd{the} SC depends on \crd{the} symmetries of the order parameter and
\crd{on} impurities. 
For example, conventional $s$-wave SCs are robust to TRS impurities because
pair breaking does not exist if the impurity concentration is small\cite{cit:JPhysChemSolids11_26}.
\crd{In contrast,} unconventional anisotropic SCs are fragile
\crd{against} TRS impurities because of anisotropic quasiparticle scattering that induce\crd{s} pair breaking\cite{cit:PRB37_4975,cit:PRB48_653,cit:PR131_1553,cit:RMP78_373}.
Such pair breaking is also induced even in $s$-wave SCs when impurities break TRS\cite{cit:JETP12_1243}.
\crd{For} the present SC, the order parameter is a mixture of $s$-wave and helical $p$-wave components.
We study quasiparticle scattering in
\crd{an} unconventional SC in order to reveal
\crd{its} similarity to or difference from \crd{those of} other SCs.

\crd{The o}rganization of this \rd{paper} is
\crd{as follows:}
In \S 2, 
in order to study the fundamental stability of the SC on \crd{the} surfaces of TIs,
we analyze the impurity concentration dependence of the mean-field order parameter in the small concentration
\crd{range}
in
\crd{an} idealistic helical Dirac electron model \crd{using} a perturbation theory referred to as the Abrikosov-Gor'kov (AG) theory\cite{cit:JETP12_1243}.
We find that such a SC\crd{,}
as well as conventional $s$-wave SCs\crd{,}
\crd{is robust}
in that the mean-field critical temperature $T_{\rm c}$ and \crd{the} mean-field order parameter $\Delta_0$ do not linearly decrease with the TRS impurity concentration\cite{cit:JPSJ80_063704}.

In \S 3, we investigate impurity effects nonperturbatively by solving the real-space Bogoliubov-de Gennes (BdG) equation for
the tight\crd{-}binding model of $\mathrm{Bi_2Se_3}$\crd{,} which is an effective model of TI\cite{cit:Nphys5_438,cit:PRB82_045122}.
We  study the scattering strength and impurity concentration dependence\crd{s} of the SC. 
Moreover\crd{,} we show that induced bound states around impurities are not Andreev bound states,
\crd{which} implies that the present SC is not a topological SC.
We summarize \crd{our study} in \S 4 with a discussion.

\section{Perturbative Approach}
\rd{In order to study impurity effects on \crd{the} surfaces of TIs
\crd{using} the AG theory, we introduce a 2D helical Dirac electron dispersion as an effective model of
SSs with an $s$-wave attractive interaction and on-site TRS scattering
following
\crd{a} previous letter.\cite{cit:JPSJ80_063704}
We introduce the $s$-wave attractive interaction because such an interaction is the most well-known origin of superconductivity, for example\crd{,}
an electron-phonon interaction.
Moreover, on-site TRS scattering
\crd{is} introduced because
\crd{it is} one of the simplest impurity scatterings and \crd{is} useful to study fundamental impurity effects.
\crd{Thus,} our Hamiltonian consists of three parts, i.e.,
a 2D helical Dirac electron dispersion, $\mathcal{H}_0$, an $s$-wave attractive interaction term, $\mathcal{H}_{\mathrm {int}}$\crd{,}
and an on-site TRS impurity scattering term, $\mathcal{H}_{\mathrm {imp}}$:
\begin{eqnarray}
\mathcal{H} = \mathcal{H}_0 + \mathcal{H}_{\mathrm{int}} + \mathcal{H}_{\mathrm{imp}}.
\end{eqnarray}}

\rd{In this section, we assume that the 2D helical Dirac electron dispersion is represented as
\begin{eqnarray}
\mathcal{H}_0 = \sum_{\bm k} c^\dagger(\bm k)[v_F(\sigma_xk_x+\sigma_yk_y)-\mu I]c(\bm k),\label{Eq2}
\end{eqnarray}
with the Fermi velocity $v_F>0$.
We set the Fermi energy $\mu>0$ in order to consider the branch of Dirac electrons above the Dirac point (called the ``+" branch \bl{hereafter}) and neglect mixture of the two branches.
Here, \crd{the} Pauli matrix $\bm\sigma$ describes \crd{the} electron
spin and $c(\bm k)=(c_{\bm k\uparrow}\ c_{\bm k\downarrow})^T$.
}

\rd{We introduce the representation in the helicity basis (helicity representation) to diagonalize $\mathcal H_0$.
By using
\crd{the} unitary transformation $d^\dagger_{\bm k,\tau} = (c^\dagger_{\bm k\uparrow} + \tau \exp(\mathrm{i}\theta_{\bm k})c^\dagger_{\bm k \downarrow})/{\sqrt{2}}\ (\tau = +\ {\rm or}\ - )$,
$\mathcal H_0$ is diagonalized as
\begin{eqnarray}
\mathcal H_0 = \sum_{\bm k} d^\dagger({\bm k}) (v_F|\bm k|\tau_z-\mu)d ({\bm k}),     \label{Eq3}
\end{eqnarray}
where $\tau_z$ is \bl{the $z$ component of \crd{the} Pauli matrix} describing branches of Dirac electrons, $\theta_{\bm k} = \arg(k_x+\mathrm{i}k_y)$, and $d(\bm k)=(d_{\bm k+}\ d_{\bm k-})^T$.
The index $\tau$ ($\tau = \pm$) represents branches of Dirac electrons.
Then, we define the energy $\xi_{\bm k}$ as the energy of the ``+" branch measured from the Fermi energy as $\xi_{\bm k} = v_F|\bm k|-\mu$.
We neglect the ``-" branch and
write $d_{+}^{(\dagger)}$ as $d^{(\dagger)}$ below.
}

\rd{\crd{Note}
that the argument
in this section is \crd{also} applicable
to the 2D helical Dirac electron dispersions with different spin-momentum relations. 
For example, in the SSs of $\mathrm{Bi_2Se_3}$, one of TIs, $\sigma_xk_y-\sigma_yk_x$ is substituted for  $\sigma_xk_x+\sigma_yk_y$ in eq. (\ref{Eq2}).\cite{cit:Nphys5_438,cit:PRB82_045122}
In
\crd{such a} case,
we have to redefine $\theta_{\bm k}$ as $\theta_{\bm k}=\arg(k_x+\mathrm{i}k_y)+\frac{\pi}{2}$.
}

\rd{We assume that
\crd{the} $s$-wave attractive interaction $H_{\rm {int}}$ is written as
\begin{eqnarray}
\mathcal{H}_{\rm int}
= 
\frac{1}{2S}\sum_{\bm k,\bm k',s,s'} V_{\mathrm{int}}(\bm k,\bm k';s,s')
c_{-\bm ks}^\dagger c_{\bm ks'}^\dagger 
c_{\bm k's'}c_{-\bm k's},
\end{eqnarray}
where  $S$ is the size of the 2D system.
In this equation, we assume that
\begin{eqnarray}
V_{\mathrm{int}}(\bm k,\bm k';\downarrow,\uparrow)=V_{\mathrm{int}}(\bm k,\bm k';\uparrow,\downarrow) =
\left\{
\begin{array}{cc}
-g&
(
\xi_{\bm k},\xi_{\bm k'}\in[-\omega_c,\omega_c])\\
0&
(
\xi_{\bm k},\xi_{\bm k'}\not\in[-\omega_c,\omega_c])
\end{array}
\right.,\nonumber\\
\end{eqnarray}
\begin{eqnarray}
V_{\mathrm{int}}(\bm k,\bm k';\uparrow,\uparrow)=V_{\mathrm{int}}(\bm k,\bm k';\downarrow,\downarrow) =
\left\{
\begin{array}{cc}
-g'&
(
\xi_{\bm k},\xi_{\bm k'}\in[-\omega_c,\omega_c])\\
0&
(
\xi_{\bm k},\xi_{\bm k'}\not\in[-\omega_c,\omega_c])
\end{array}
\right.,\nonumber\\
\end{eqnarray}
with
\crd{a} cutoff $\omega_c  \ll \mu$
and $g, g' > 0$.
\crd{Note} that,
\crd{under} the condition $\omega_c \ll \mu$,
the interaction
affects \crd{only} electrons on the ``+" branch.
Then the interaction term in the helicity representation is
\begin{eqnarray}
\mathcal H_{\mathrm {int}} \simeq
-\frac{g}{4S}\sum_{\bm k,\bm k'}
e^{\mathrm{i}(\theta_{\bm k'}-\theta_{\bm k})}d^\dagger_{-\bm k}d^\dagger_{\bm k}d_{\bm k'}d_{-\bm k'} . \label{2.7}
\end{eqnarray}}

\rd{Here\crd{,} we introduce an on-site TRS impurity scattering term as
\begin{eqnarray}
\mathcal{H}_{\rm imp}
=
\frac{u}{S}\sum_{i=1}^{N_{\rm i}}\sum_{\bm k,\bm k'}e^{\mathrm{i}(\bm k'-\bm k)\cdot\bm R_i} 
c^\dagger(\bm k)c({\bm k'}), \label{Eq8}
\end{eqnarray}
where
$N_{\rm i}$ is the number of the impurities in the system
and $\bm R_i\ (i = 1,\cdots,N_{\rm i})$ is the impurity location.
Then the impurity scattering term in the helicity representation is
\begin{eqnarray}
H_{\rm imp}
=
\frac{u}{S}
\sum_{i=1}^{N_{\rm i}}
\sum_{\bm k,\bm q}e^{-\mathrm{i}\bm q\cdot\bm R_i}
P(\theta_{\bm k}-\theta_{\bm k+\bm q})
d^\dagger_{\bm k+\bm q} d_{\bm k},
\end{eqnarray}
where $P(\theta)=\exp(\mathrm{i}\theta/2)\cos(\theta/2)$ is a phase factor specific to Dirac electron systems.
Here, $P(\pi)=0$ means that the backscattering is forbidden.
Moreover, this phase factor contributes to the $\pi$ Berry phase, which leads to
\crd{an} antilocalization effect of single Dirac cone systems\cite{cit:JPSJ67_2857}, i.e., the electric conductivity in the system
has \crd{a} positive quantum correction.}

\rd{We introduce a mean-field approximation in eq. (\ref{2.7}) and construct
\crd{a} BCS mean-field Hamiltonian.
By the BCS\crd{-}type decoupling, a momentum-dependent pair potential $\Delta(\bm k)$ is derived as
\begin{eqnarray}
\Delta(\bm k) 
= \frac{g}{2S}e^{-\mathrm{i}\theta_{\bm k}}\sum_{\bm k'} e^{\mathrm{i}\theta_{\bm k'}}\langle d_{\bm k'}d_{-\bm k'}\rangle = \Delta e^{-\mathrm{i}\theta_{\bm k}}\label{PairPotential},
\end{eqnarray}
and our interaction term is approximated as 
\begin{eqnarray}
\mathcal{H}_{\rm int} \simeq -\frac{1}{2}\sum_{\bm k} \left[\Delta(\bm k)d^\dagger_{-\bm k}d^\dagger_{\bm k} + \Delta^*(\bm k)d_{\bm k}d_{-\bm k} \right].\label{2.11}
\end{eqnarray}
The pair potential in eq. (\ref{PairPotential}) resembles a spinless chiral $p$-wave pair potential.
This resemblance is related to the emergence of Majorana bound states around integer vortices\cite{cit:PRL100_096407,cit:PRB61_10267}.
The difference between the two SCs is that the SC of helical Dirac electrons is time\crd{-}reversal\crd{-}symmetric\crd{,}
but \crd{the} spinless $p$-wave SC is not.}

\rd{In the representation of the original electrons operator $c$,
this SC is composed of a mixture of $s$-wave (singlet) and $p$-wave (triplet) symmetr\crd{ies}, and the interaction term is represented as
\begin{eqnarray}
\mathcal{H}_{\rm int} = -\sum_{\bm k} \sum_{s_1,s_2} \left[c_{-\bm ks_1}^\dagger \hat\Delta_{s_1s_2}(\bm k) c_{\bm ks_2}^\dagger + h.c. \right],
\end{eqnarray}
where
\begin{eqnarray}
\hat\Delta(\bm k)
\nonumber&=&
\hat\Delta_{s}(\bm k)+\hat\Delta_t(\bm k)\\
\hat\Delta_{s}(\bm k) &=& \frac{\Delta}{2}
\begin{pmatrix}
0 & 1 \\ -1 & 0
\end{pmatrix}
,\nonumber\\
\hat\Delta_t(\bm k)&=&\frac{\Delta}{2}
\begin{pmatrix}
e^{-\mathrm{i}\theta_{\bm k}} & 0 \\ 0 & -e^{\mathrm{i}\theta_{\bm k}}
\end{pmatrix}.
\end{eqnarray}
Here, \crd{the} $s$-wave pairing $\hat\Delta_s$ and \crd{the} $p$-wave pairing $\hat\Delta_t$ are introduced. 
The mixture of two components
\crd{originates} from the broken inversion symmetry in the surface.
The $d$ vector corresponding to the triplet pairing component $\hat\Delta_t$ has a momentum dependence $\bm d(\bm k) \propto (k_x,k_y,0)$, i.e., $\hat\Delta_t$ has a helical $p$-wave symmetry.
Here, the helical $p$-wave SC is TRS $p$-wave SC, in which the chirality of the pair potential is different for each spin component.\cite{cit:PRL102_187001}}

By using the mean-field approximation \rd{in eq. (\ref{2.11})}, the Hamiltonian is approximated as
\rd{\begin{eqnarray}
\mathcal H 
&\simeq&\nonumber
\frac{1}{2} \Psi^\dagger\hat H\Psi \\
&=& \nonumber 
\frac{1}{2} \Psi^\dagger\left(\hat H_{\mathrm{MF}}+\hat V_\mathrm{imp} \right)\Psi  \\
&=&\nonumber
\frac{1}{2}\sum_{\bm k}\Psi^\dagger(\bm k)\hat H_{\mathrm{MF}}(\bm k)\Psi(\bm k)\\
&+&\nonumber
\frac{1}{2}\frac{u}{S}\sum_{i=1}^{N_{\mathrm{i}}}\sum_{\bm k,\bm k'}e^{\mathrm i(\bm k'-\bm k)\cdot\bm R_i}\Psi^\dagger(\bm k)\hat V_1(\bm k,\bm k')\Psi(\bm k')\label{BCS_MF}.\\
\end{eqnarray}}
We introduce \crd{a} Nambu representation
$
\Psi(\bm k)= ( \Psi_{\bm k,e} \ \Psi_{\bm k,h} )^T 
\equiv( d_{\bm k}\ d^\dagger_{-\bm k})^T.
$
\crd{T}he matrix elements in eq. (\ref{BCS_MF}) are defined as
\begin{eqnarray}
\hat H_{\mathrm{MF}}(\bm k) &=&\nonumber
\begin{pmatrix}\xi_{\bm k} &\Delta e^{-\mathrm i\theta_{\bm k}} \\ \Delta^*e^{\mathrm i\theta_{\bm k}} & -\xi_{\bm k} \end{pmatrix},\\
\hat V_1(\bm k,\bm k') &=& \begin{pmatrix} P(\theta_{\bm k'}-\theta_{\bm k}) & 0 \\ 0 & -P(\theta_{-\bm k}-\theta_{-\bm k'}) \end{pmatrix}.\nonumber \\
\end{eqnarray}

We define an imaginary time Green's function as $\hat G(\tau) = -\left\langle T_\tau \Psi(\tau)\Psi^\dagger(0) \right\rangle$,
where the time evolution of the operators $\Psi^{(\dagger)}$
\crd{is} obtained from $\Psi^{(\dagger)}(\tau)=e^{\tau\mathcal H}\Psi^{(\dagger)} e^{-\tau\mathcal H}$.
Then the Green's function in frequency space is $\hat G(\mathrm i\omega_n) = \int^\beta_0 d\tau\ e^{\mathrm i\omega_n\tau} \hat G(\tau)$.
From the equation of motion of the Green's function, \crd{the} Gor'kov equation 
\begin{eqnarray}
(\mathrm i\omega_n-\hat H)\hat G(\mathrm i\omega_n) = \hat 1
\end{eqnarray}
is derived, where $\omega_n = (2n+1)\pi T$ is the fermionic Matsubara frequency
and $T$ is the temperature. 

When we assume that the scattering term $\hat V_\mathrm{imp}$ is perturbation
using a perturbation series expansion with respect to $u/S$, the Green's function $\hat G$ is represented as
\begin{eqnarray}
\hat G(i\omega_n) =\hat G_0(i\omega_n) +\sum_{n=1}^{\infty}  (\hat G_0(i\omega_n) \hat V_\mathrm{imp})^n\hat G_0(i\omega_n)  \label{pert_series},\nonumber\\
\end{eqnarray}
where the nonperturbative Green's function $\hat G_0(\mathrm i\omega_n) = (\mathrm i\omega_n-\hat H_{\mathrm{MF}})^{-1}$ is introduced.

According to the AG theory, we perform the impurity average operation so that the system recovers \crd{its} translational symmetry.
By this operation, diagonal terms of momentum remain in the Hamiltonian and we only have to consider diagonal terms in the Green's function.
By the impurity average operation, the quantity
$
\frac{1}{S}\sum_{i,j=1}^{N_{\rm i}} e^{\mathrm i(\bm q\cdot\bm R_i-\bm q'\cdot\bm R_j)}
$
is replaced
\crd{with} 
$
n_\mathrm{i}\delta_{\bm q,\bm q'}\crd{,}
$
where $n_\mathrm{i}=N_\mathrm{i}/S$ is the impurity concentration.
We perform the impurity average of the right\crd{-}hand side of eq. (\ref{pert_series}) for each term.
The terms
\crd{that} correspond to $n=1$ and $n=2$ in eq. (\ref{pert_series}) are represented as diagrams in Fig\crd{s}. 1(a) and
1(b)\crd{, respectively}.
They are the lowest\crd{-}order terms about $n_\mathrm{i}$.
However, the term shown in Fig. 1(a) is negligible for the estimation of $T_{\rm c}$ because
\crd{it} just causes a constant self-energy shift and does not contribute to relaxation processes due to
pair breaking.

\begin{figure}[htbp]
\begin{center}
\includegraphics[width=80mm]{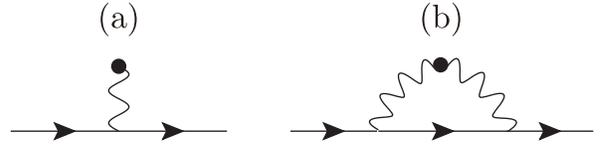}
\caption{Diagrams for
self-energy terms of the lowest order in eq. (\ref{pert_series}). (a) and (b) correspond to $n=1$ and $n=2$, respectively.
Wavy lines represent impurity scatterings.}
\end{center}
\end{figure}

Therefore\crd{,} we consider the diagram \crd{in} Fig. 1(b) and its higher\crd{-}order series as the self-energy term for the estimation of $T_{\mathrm c}$.
The contribution of this term is $\mathcal{O}(n_{\mathrm i})$.
We adopt this approximation following Abrikosov-Gor'kov\cite{cit:JETP12_1243} and Ambegaokar-Griffin\cite{cit:PR137_A1151}.
Then the Green's function is calculated as
\begin{eqnarray}
\hat G(\bm k,\mathrm i\omega_n) &= &\nonumber
\frac{1}{\left[\hat G_0(\bm k,\mathrm i\omega_n)\right]^{-1}-\hat\Sigma(\bm k,\mathrm i\omega_n)}\\
 &=& 
\frac{1}{\mathrm i\tilde\omega_n-\hat{\tilde H}_{\mathrm{MF}}(\bm k,\mathrm i\omega_n)},
\label{renormalized_G}
\end{eqnarray}
where the self-energy term is 
\begin{eqnarray}
\hat\Sigma(\bm k,\mathrm i\omega_n) 
&=&\nonumber \frac{n_\mathrm{i}u^2}{S}\sum_{\bm k'}\hat V_1(\bm k,\bm k')\hat G(\bm k',\mathrm i\omega_n) \hat V_1(\bm k',\bm k)\\
&=&\nonumber -\frac{n_\mathrm{i}u^2}{S}\sum_{\bm k'}\frac{\left|P(\theta_{\bm k'}-\theta_{\bm k})\right|^2}{\tilde\omega_n^2+|\tilde\Delta_n|^2+\xi_{\bm k'}^2}\\
&\times&
\begin{pmatrix}
\mathrm i\tilde\omega_n+\xi_{\bm k'} &-\tilde\Delta_n e^{-\mathrm i\theta_{\bm k}}\\
-\tilde\Delta_n^* e^{\mathrm i\theta_{\bm k}} & \mathrm i\tilde\omega_n-\xi_{\bm k'} \\
\end{pmatrix},
\label{self_energy}
\end{eqnarray}
and the renormalized mean-field Hamiltonian is
\begin{eqnarray}
\hat{\tilde H}_{\mathrm{MF}}(\bm k,\mathrm i\omega_n) =
\begin{pmatrix}\xi_{\bm k} & \tilde\Delta_ne^{-\mathrm i\theta_{\bm k}} \\ \tilde\Delta_n^*e^{\mathrm i\theta_{\bm k}} & -\xi_{\bm k} \end{pmatrix}. \label{renormalized_H}
\end{eqnarray}
In eq\crd{s}. (\ref{renormalized_G})
\crd{-}(\ref{renormalized_H}), $\tilde\omega_n$ and $\tilde\Delta_n$ are the renormalized frequency and pair potential, respectively.

By using eq\crd{s}. (\ref{renormalized_G})
\crd{-}(\ref{renormalized_H}), $\tilde\omega_n$ and $\tilde\Delta_n$ are self-consistently calculated as
\begin{eqnarray}
\tilde\omega_n &=& \omega_n  + \frac{1}{2\tau_1}\frac{\tilde\omega_n}{ \sqrt{\tilde \omega_n^2 + |\tilde\Delta_n|^2} } \label{renormalized_freq},\\ 
\tilde\Delta_n &=& \Delta + \frac{1}{2\tau_2}\frac{\tilde\Delta_n}{\sqrt{\tilde\omega_n^2+  |\tilde\Delta_n|^2}  },\label{renormalized_pair}
\end{eqnarray}
where $\tau_1$ and $\tau_2$ are two types of relaxation times.
In the case of the present SC, the two relaxation times are calculated as
\begin{eqnarray}
\frac{1}{2\tau_1}=\frac{1}{2\tau_2}&=&\nonumber
 \pi n_\mathrm{i} u^2N_0  \int^{2\pi}_0\frac{d\theta}{2\pi}\ \cos^2\left(\theta/2\right)\\
&=& \frac{\pi n_\mathrm{i} u^2 N_0}{2},
\end{eqnarray}
where  $N_0 = \frac{1}{S}\sum_{\bm k}\delta(\xi_{\bm k})$ is the density of states (DOS) at the Fermi energy.
Therefore $\tau_1$ and $\tau_2$ are the same
\crd{when} considering $\mathcal{O}(n_{\mathrm i})$. 

In order to estimate the order parameter or
critical temperature, we need
\crd{a} self-consistent equation.
\crd{A} self-consistent equation is obtained
\crd{using} eq. (\ref{PairPotential})\crd{:}
\begin{eqnarray}
\Delta =
 -\frac{g}{2\beta S}\sum_{\mathrm i\omega_n}\sum_{\bm k}e^{\mathrm i\theta_{\bm k}}G_{e,h}(\bm k,\mathrm i\omega_n).\label{self consistent}
\end{eqnarray}

The self-consistent equation for the order parameter at zero temperature\crd{,} $\Delta_0$\crd{,} is
\begin{eqnarray}
\Delta_0(n_{\rm i}) = \frac{g}{2\beta S}\int^{\omega_c}_{-\omega_c}\frac{d\omega}{2\pi}\sum_{\bm k}
\frac{\tilde\Delta}{\tilde\omega^2 + \xi_{\bm k}^2+\tilde\Delta^2}\label{zerotemp},
\end{eqnarray}
where $\tilde\omega = \omega  + \frac{1}{2\tau_1}\frac{\tilde\omega}{ \sqrt{\tilde \omega^2 + |\tilde\Delta|^2} }$ and $\tilde\Delta = \Delta + \frac{1}{2\tau_2}\frac{\tilde\Delta}{\sqrt{\tilde\omega^2+  |\tilde\Delta|^2}  }$.

On the other hand, the self-consistent equation for the mean-field critical temperature $T_\mathrm{c}$ is
\begin{eqnarray}
1=\frac{gT_{\mathrm c}(n_\mathrm i)}{2S}\sum_{\mathrm i\omega_n}\sum_{\bm k}
\frac{(2\tau_2|\tilde\omega_n|)^{-1}}{\tilde\omega_n^2 + \xi_{\bm k}^2} \label{criticaltemp}.
\end{eqnarray}
Because the long-range SC order does not develop in 2D systems at \crd{a} finite temperature,
$T_{\rm c}$ calculated from the mean-field theory provides a criterion of the Berezinskii-Kosterlitz-Thouless (BKT) transition for
\crd{the} development of
\crd{a} quasi-long-range order\cite{cit:JETP34_610,cit:JPhysC6_1181}.

By \crd{using the above}
equations, the values are obtained as
\begin{eqnarray}
T_{\mathrm c}(n_{\mathrm{i}}) &=& T_{\mathrm c}(0) - \frac{\pi}{4\tau_s(n_{\mathrm{i}})},\ \  T_c(0)=\frac{2e^\gamma}{\pi}\omega_c\exp\left(-\frac{2}{gN_0}\right),\nonumber \\ \\
\Delta_0(n_{\mathrm{i}}) &=& \Delta_0(0) - \frac{\pi}{4\tau_s(n_{\mathrm{i}})},\ \  \Delta_0(0)=2\omega_c\exp\left(-\frac{2}{gN_0}\right)\crd{,} \nonumber\\ 
\end{eqnarray}
where the relaxation time $\tau_s$ is defined by 
\begin{eqnarray}(\tau_s)^{-1} = (2\tau_1)^{-1}-(2\tau_2)^{-1}.
\end{eqnarray}
%
Because $(\tau_s)^{-1} = \mathcal{O}(n_\mathrm{i}^2)$ is satisfied, $T_\mathrm{c}$ and $\Delta_0$ do not decrease linearly in the present SC.

We compare this result with the results for other full-gap SCs
\crd{reported} in the literature\cite{cit:RMP78_373,cit:JETP12_1243}\crd{.}
Table I shows the results of the present SC in comparison with other 2D SCs.
For each SC, $(2\tau_1)^{-1}$ and $(2\tau_2)^{-1}$ are calculated.
We call SCs with $\tau_s^{-1} = \mathcal{O}(n_{\mathrm i}^2)$
stable SCs, while SCs \crd{with $(\tau_s)^{-1}=\mathcal{O}(n_{\mathrm i})$}
are fragile\crd{.}
In the table, ``magnetic scattering"
\crd{indicates} that the impurity Hamiltonian takes a form as
\begin{eqnarray}
\mathcal{H}_{\mathrm{magimp}} = \frac{u}{S}\sum_{i=1}^{N_{\rm i}}c^\dagger_{n(i)}\sigma_z c_{n(i)},
\end{eqnarray}
\crd{which} represents the scattering by magnetic impurities polarized along the $z$\crd{-}direction.
The forms of matrix elements for the BCS mean-field Hamiltonian $\hat H_{\mathrm{MF}}(\bm k)$ and the scattering Hamiltonian $\hat V_1(\bm k,\bm k')$ are also shown in the table.
\onecolumn
\begin{table}[h]
\caption{Stability of SCs:
(a) $S$-wave SC with TRS impurities.
(b) $S$-wave SC with magnetic impurities.
(c) Chiral $p$-wave SC with TRS impurities. (d) SC on TIs with TRS impurities. (e) SC on TIs with magnetic impurities.}
\begin{center}
\begin{tabular}{|c|c|c|c|}
\hline                                        & $\hat H_{\mathrm{MF}}(\bm k)$  & $\hat V_1(\bm k,\bm k')$ & stability \\
\hline  (a)  & $\begin{pmatrix}\xi(\bm k)&\Delta\\ \Delta^*&-\xi_{\bm k}\end{pmatrix}$& $\begin{pmatrix}1&0\\0&-1\end{pmatrix}$ & stable  \\
\hline  (b)   & $\begin{pmatrix}\xi(\bm k)&\Delta\\ \Delta^*&-\xi_{\bm k}\end{pmatrix}$& $\begin{pmatrix}1&0\\0&1\end{pmatrix}$ & fragile \\
\hline  (c) & $\begin{pmatrix}\xi(\bm k)&\Delta e^{-i\theta_{\bm k}}\\ \Delta^*e^{i\theta_{\bm k}}&-\xi_{\bm k}\end{pmatrix}$& $\begin{pmatrix}1&0\\0&-1\end{pmatrix}$ & fragile\\
\hline  (d) & $\begin{pmatrix}\xi(\bm k)&\Delta e^{-i\theta_{\bm k}}\\ \Delta^*e^{i\theta_{\bm k}}&-\xi_{\bm k}\end{pmatrix}$& $\begin{pmatrix}P(\theta_{\bm k'}-\theta_{\bm k})&0\\0&-P(\theta_{-\bm k}-\theta_{-\bm k'})\end{pmatrix}$ & stable\\
\hline  (e) & $\begin{pmatrix}\xi(\bm k)&\Delta e^{-i\theta_{\bm k}}\\ \Delta^*e^{i\theta_{\bm k}}&-\xi_{\bm k}\end{pmatrix}$& $\begin{pmatrix}P(\theta_{\bm k'}-\theta_{\bm k}+\pi)&0\\0&-P(\theta_{-\bm k}-\theta_{-\bm k'}+\pi)\end{pmatrix}$ &fragile\\
\hline
\end{tabular}
\end{center}
\end{table}
\twocolumn

To summarize this section,
unconventional SCs induced by the $s$-wave attractive interaction on \crd{the} surfaces of TIs are robust to TRS impurities.
This result is achieved by calculating the dependences of $T_{\rm c}$ and $\Delta_0$ on the TRS impurity concentration, where $T_{\rm c}$ provides
a criterion of the BKT transition for
\crd{a} quasi-long-range order in 2D systems.
The unconventional SC on the surface of TI is robust because of the cancellation of two phase factors, one from the pairing potential and the other arising when a Dirac electron is scattered by a TRS impurity.
\crd{In contrast,} unconventional SCs
\crd{reported} in the literature, such as the $d$-wave and chiral $p$-wave SC\crd{s}\cite{cit:PRB37_4975,cit:PRB48_653,cit:PR131_1553,cit:RMP78_373},
are sensitively suppressed through scattering
by a tiny concentration of impurities because of the phase factor of the pairing potential.

We treated impurities as perturbations in this section.
The perturbation theory is valid if
$u^2n_\mathrm{i}N_0$ is much smaller than the pair potential $\Delta$ \crd{though}
a rough estimate.
This estimation is derived from the reduction
\crd{in} the pair potential due to magnetic scattering\crd{.}

\section{Nonperturbative Approach}

\subsection{Model and \crd{m}ethod}
In order to study \crd{the} impurity effects on \crd{the} SC on \crd{the} surfaces of TIs by
real-space BdG calculation, we use a tight\crd{-}binding model of $\mathrm{Bi_2Se_3}$ in
slab geometry, which is an effective model of 3D TI\cite{cit:Nphys5_438,cit:PRB82_045122}.
We obtain a single Dirac cone
\crd{on} the surface of the slab.

In order to study impurity effects, we consider a Hamiltonian composed of three terms \rd{in the same way as} \crd{in} \S 2\crd{:}
\begin{eqnarray}
\mathcal{H} = \mathcal{H}_0 + \mathcal{H}_{\mathrm{int}} + \mathcal{H}_{\mathrm{imp}}.
\end{eqnarray}
Here, $\mathcal H_0$ is the effective tight-binding Hamiltonian of $\mathrm{Bi_2Se_3}$ , which is first introduced by Zhang {\em et al}.\cite{cit:Nphys5_438}
On the other hand, $\mathcal{H}_{\mathrm{int}}$ is the interaction term and $\mathcal{H}_{\mathrm{imp}}$ is the \rd{impurity potential}.

The Hamiltonian $\mathcal{H}_0$ has a structure of a $4\times4$ matrix, because of the presence of two orbitals and spin indices.
\crd{The t}wo orbitals are the antibonding and bonding orbitals constructed from the $p_z$-orbitals of Bi and Se atoms\crd{;}
they have different parities \crd{from} each other.
We refer to these orbitals as ``$E$'' and ``$H$'' orbitals, respectively.
$\mathrm{Bi_2Se_3}$ has a rhombohedral symmetry, but for simplicity we adopt a model reduced to the $\mathrm{D_{4h}}$ symmetry\cite{cit:Nphys5_438}.
In studying fundamental properties of
SSs near the Dirac point, this approximation is valid since anisotropy owing to the rhombohedral symmetry is weak near the Dirac point.

A Bloch representation of $\mathcal{H}_0$\cite{cit:Nphys5_438} is
\begin{eqnarray}
\small
&&\nonumber\hat H_0(\bm k)\nonumber\\&=&
\begin{pmatrix}
\epsilon_{\bm k}+M_{\bm k} &A_zk_z &0 &A_{||}k_{-} \\
A_zk_z &\epsilon_{\bm k}-M_{\bm k} & A_{||} k_-&0 \\
0 &A_{||} k_{+} &\epsilon_{\bm k}+M_{\bm k} &-A_zk_z \\
A_{||}k_{+} &0 &-A_zk_z &\epsilon_{\bm k}-M_{\bm k} 
\end{pmatrix},\label{Bi2Se3_Model}
\end{eqnarray}
where
\begin{eqnarray*}
\mathcal H_0 = \sum_{\bm k}c^\dagger(\bm k)\hat H_0(\bm k) c(\bm k),\ 
c(\bm k)=
\begin{pmatrix}c_{\bm k\uparrow \mathrm{E}}&c_{\bm k\uparrow \mathrm H}&c_{\bm k\downarrow \mathrm{E}}&c_{\bm k\downarrow \mathrm H} \end{pmatrix}^T,
\end{eqnarray*}
\begin{eqnarray*}
\epsilon_{\bm k} &=& D_{||}k_{||}^2 +D_{z}k_z^2, \\
M_{\bm k} &=& M-B_{||}k_{||}^2-B_zk_z^2\ \ \ (B_{||},B_z<0),
\end{eqnarray*}
\begin{eqnarray*}
k_{\pm}=k_x\pm \mathrm{i}k_y, \bm k_{||}=(k_x,k_y),
\end{eqnarray*}
where $A_{||}$ and $A_z$ represent the strength of SOIs, while $M$ is the energy difference between \rd{the} two orbitals.
The band curvatures of the two subbands are different,
\crd{which} arises from nonzero $B_{||}$ and $B_z$.
Here\crd{,} we note that, due to
\crd{the} gauge transformation defined in eq. (18) of ref. \citen{cit:PRB82_045122},
$\hat{H}_{0}({\bm k})$ is not invariant under \crd{the} operation of a standard choice for the time reversal operator, $I\otimes i \sigma_{y} \cdot \mathcal{K}$,
where $I$ is the identity matrix acting on subband indices, $\sigma_{y}$ is the Pauli matrix acting on the spin indices, and $\mathcal{K}$ is the complex conjugate operator.

In order to construct a Wannier representation of $\mathcal{H}_0$, we perform a substitution such that
$$
k_i \rightarrow \sin k_i, k_i^2 \rightarrow 2-2\cos k_i\ \ \ (i=x,y,z),
$$
and
\crd{the} Fourier transformation
\begin{eqnarray}
c(\bm k) &=& \frac{1}{\sqrt{N_xN_yN_z}}\sum_{x=1}^{N_x}\sum_{y=1}^{N_y}\sum_{z=1}^{N_z}e^{\mathrm i\bm k\cdot\bm r}c(\bm r),\\
c^\dagger(\bm k) &=& \frac{1}{\sqrt{N_xN_yN_z}}\sum_{x=1}^{N_x}\sum_{y=1}^{N_y}\sum_{z=1}^{N_z}e^{-\mathrm i\bm k\cdot\bm r}c^\dagger(\bm r).
\end{eqnarray}

For simplicity\crd{,} we choose the model parameters in eq. (\ref{Bi2Se3_Model}) as
\rd{\begin{eqnarray}
A_x&=&1\ \mathrm{eV},A_z=0.5\ \mathrm{eV},B_x=1.5\ \mathrm{eV},B_z=0.3\ \mathrm{eV}, \nonumber\\
D_x&=&D_z=0,M=0.5\ \mathrm{eV},
\end{eqnarray}}
where the particle-hole symmetry about subbands $E$ and $H$ is imposed.
\rd{We assume that the lattice parameter is normalized as $1$.}
\textcolor{black}{
Then the curvatures of two subbands are equivalent except signs. For readers, we cite the parameters given by}
{\em{ab initio}} calculations\cite{cit:Nphys5_438} below: 
\rd{\begin{eqnarray}
A_x&=&0.5\ \mathrm{eV},A_z=0.2\ \mathrm{eV},B_x=0.6\ \mathrm{eV},B_z=0.1\ \mathrm{eV}, \nonumber\\
D_x&=&0.1\ \mathrm{eV}, D_z=0.05\ \mathrm{eV},M=0.3\ \mathrm{eV}.
\end{eqnarray}}

We obtain a single Dirac cone at the $\Gamma$ point in a surface by considering a slab geometry when the parameters support nontrivial $Z_2$ topological indices.
In this paper, we impose an open boundary condition in \crd{the} $z$\crd{-}direction and periodic boundary conditions in \crd{the} $x$\crd{-} and $y$\crd{-}directions.
Figure 2 shows \bl{energy dispersions} for $\mathcal H_0$.
In the dispersion, gapless SSs exist inside the bulk band gap.

\begin{figure}[htbp]
\begin{center}
\includegraphics[width=80mm]{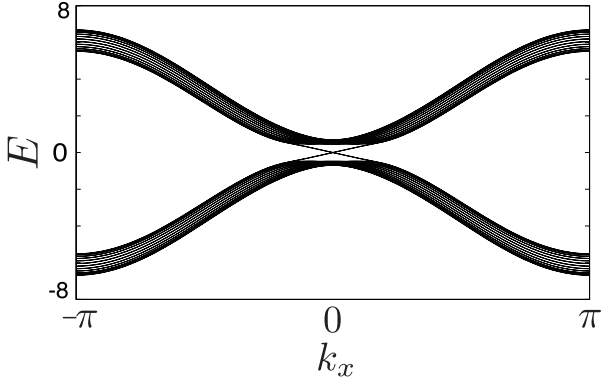}
\caption{Dispersion when
\crd{the} set of Hamiltonian parameters is \rd{$A_x=1\ \mathrm{eV},A_z=0.5\ \mathrm{eV},B_x=1.5\ \mathrm{eV},B_z=0.3\ \mathrm{eV},D_x=D_z=0 \mathrm{\ and\ } M=0.5\ \mathrm{eV}$}.
An open\crd{-}boundary condition is imposed in the $z$\crd{-}direction and $N_z=12$. \rd{We show the case} \crd{of} $k_y=0$.}
\label{Energy dispersion}
\end{center}
\end{figure}

Then, we introduce
\crd{the} $s$-wave attractive interaction term $\mathcal{H}_\mathrm{int}$ as 
\begin{eqnarray}
\mathcal H_{\mathrm{int}} 
= -\frac{1}{2} \sum_{i,j,\sigma,\sigma',\tau,\tau'}c^\dagger_{i\tau\sigma}c^\dagger_{j\tau'\sigma'}g_{\sigma\sigma'}^{\tau\tau'}(i,j)c_{j\tau'\sigma'}c_{i\tau\sigma}\label{Ineraction},
\end{eqnarray}
where $g_{\sigma\sigma'}^{\tau\tau'}(i,j) = g_{\sigma'\sigma}^{\tau'\tau}(j,i)$ holds.

As the third term of our model Hamiltonian, we introduce
\crd{the} $s$-wave TRS impurity scattering Hamiltonian $\mathcal{H}_\mathrm{imp}$.
The impurity scattering Hamiltonian $\mathcal{H}_\mathrm{imp}$ is represented as
\begin{eqnarray}
\mathcal{H}_{\mathrm{imp}} = u\sum_{i=1}^{N_{\mathrm{i}}}c^\dagger_{n(i)}c_{n(i)}= c^\dagger\hat H_{\mathrm{imp}}c,
\end{eqnarray}
where $N_{\mathrm i}$ is the number of impurities, $n(i)$ is the location of the $i$-th impurity, $u$ is the strength of the impurity potential\crd{,}
and $\hat H_{\rm imp}$ is the matrix representation of the Hamiltonian.

By solving the BdG equation for the introduced model Hamiltonian, we analyze the impurity effects on the SC on \crd{the} surfaces of TIs.
\bl{To construct the BdG equation, we perform
\crd{the} mean-field approximation}
\begin{eqnarray}&&\nonumber
c^\dagger_{i\tau\sigma}c^\dagger_{j\tau'\sigma'}c_{j\tau'\sigma'}c_{i\tau\sigma}\\
&\rightarrow&\nonumber
\langle c^\dagger_{i\tau\sigma}c^\dagger_{j\tau'\sigma'}\rangle c_{j\tau'\sigma'}c_{i\tau\sigma}
+
c^\dagger_{i\tau\sigma}c^\dagger_{j\tau'\sigma'} \langle c_{j\tau'\sigma'}c_{i\tau\sigma}\rangle\\
&-&
\langle c^\dagger_{i\tau\sigma}c^\dagger_{j\tau'\sigma'} \rangle\langle c_{j\tau'\sigma'}c_{i\tau\sigma}\rangle\rd{,}\label{3.11}
\end{eqnarray}
\rd{which leads to the BCS mean-field Hamiltonian}
\begin{eqnarray}
\mathcal H \simeq \frac{1}{2}\Psi^\dagger\hat H_{\mathrm{MF}}\Psi,
\end{eqnarray}
where $\Psi=(\Psi_{e}\ \Psi_{h})^T=(c\ [c^\dagger]^T )^T$ is a Nambu spinor
and
\begin{eqnarray}
\hat H_{\mathrm{MF}} &=&\nonumber 
\begin{pmatrix}
\hat H_0+\hat H_{\mathrm{imp}} +\mu \hat I & \hat\Delta\\
\hat\Delta^\dagger  & -[\hat H_0+\hat H_{\mathrm{imp}}  + \mu \hat I]^T
\end{pmatrix}\\
&+&
E_\Delta
\begin{pmatrix}
 \hat I & 0 \\
0 &  \hat I
\end{pmatrix}\label{3.19}
\end{eqnarray}
is the matrix element\crd{.}
\crd{H}ere\crd{,} $\mu$ is the Fermi energy, $E_\Delta$ is a constant energy shift\crd{,} and $\hat I$ is the identity matrix.
The constant energy shift $E_\Delta$
\crd{originates} from the terms like $\langle cc\rangle\langle c^\dagger c^\dagger\rangle$ in eq. (\ref{3.11}) and calculated as $E_{\Delta} = \sum_{ij\sigma\sigma'\tau\tau'} g^{\tau\tau'}_{\sigma\sigma'}\langle c^\dagger_{i\tau\sigma}c^\dagger_{j\tau'\sigma'} \rangle\langle c_{j\tau'\sigma'}c_{i\tau\sigma}\rangle.$
Since \crd{the} contribution of this term is just a constant energy shift, we are able to neglect this term in determining the mean-field self-consistently.
In eq. (\ref{3.19}), the pair potential is introduced as
\begin{eqnarray}
\Delta^{\tau\tau'}_{\sigma\sigma'}(i,j)=g^{\tau\tau'}_{\sigma\sigma'}(i,j)\langle c_{i\tau\sigma}c_{j\tau'\sigma'} \rangle.\label{3.12}
\end{eqnarray}
The $4\times4$ pair potential matrix $\hat\Delta$ is expressed as
\begin{eqnarray}
\hat\Delta = 
\begin{pmatrix}
\Delta^{EE}_{\uparrow\uparrow} & \Delta^{EH}_{\uparrow\uparrow} & \Delta^{EE}_{\uparrow\downarrow} & \Delta^{EH}_{\uparrow\downarrow} \\
\Delta^{HE}_{\uparrow\uparrow} & \Delta^{HH}_{\uparrow\uparrow} & \Delta^{HE}_{\uparrow\downarrow} & \Delta^{EH}_{\uparrow\downarrow} \\
\Delta^{EE}_{\downarrow\uparrow} & \Delta^{EH}_{\downarrow\uparrow} & \Delta^{EE}_{\downarrow\downarrow} & \Delta^{EH}_{\downarrow\downarrow} \\
\Delta^{HE}_{\downarrow\uparrow} & \Delta^{HH}_{\downarrow\uparrow} & \Delta^{HE}_{\downarrow\downarrow} & \Delta^{HH}_{\downarrow\downarrow} 
\end{pmatrix}.
\end{eqnarray}
\crd{Note} that $\Delta^{\tau\tau'}_{\sigma\sigma'}(i,j)=-\Delta^{\tau'\tau}_{\sigma'\sigma}(j,i)$ is satisfied because of the Pauli exclusion principle.

By diagonalizing $\hat H_{\mathrm{MF}}$,
we obtain the excitation energy spectrum of the Bogoliubov quasiparticles $E_\nu$ and the set of eigenvectors $w_{\nu}(i\tau\sigma r)$ corresponding to amplitudes of the quasiparticle wave functions,
that is\crd{,}
\begin{eqnarray}
\hat H_{\mathrm{MF}} w_\nu = E_\nu w_\nu.
\end{eqnarray}
The indices $i$, $\tau$, $\sigma$, and $r$ represent the site, subbands, spin, and
particle-hole indices, respectively.

The mean-field Hamiltonian has the particle-hole symmetry (PHS), i.e.
\begin{eqnarray}
\{\hat C, \hat H_{\mathrm{MF}} \}=0, 
\end{eqnarray}
where $\hat C=\hat r_x \hat K$ and $\hat r_x$ transforms a particle and a hole each other by the Pauli operator in the particle hole space and $\hat K$ is the complex conjugate operator.
Then $w_\nu$ and $\hat Cw_\nu$ are referred to as a particle-hole pair, i.e.,  $\hat H_\mathrm{MF} \hat Cw_\nu = -E_\nu \hat Cw_\nu$ is satisfied.
For convenience, we redefine the index $\nu$ of an eigenvalue so that $w_\nu$ and $w_{-\nu}$ are a particle-hole pair, where $\nu>0$ \crd{and} $E_\nu>0$.
Then, $w_{-\nu} = \hat Cw_{\nu}$ and $E_{-\nu}=-E_\nu$ hold.

Then we introduce \crd{the} creation and annihilation operators of
quasiparticles, $\alpha_\nu$.
Here, $\Psi$ and $\alpha_\nu$ are related by \crd{the} Bogoliubov transformation
\begin{eqnarray}
\Psi =\sum_\nu w_\nu \alpha_\nu= \sum_{\nu>0} \left(w_\nu \alpha_\nu  + w_{-\nu}\alpha_\nu^\dagger \right), 
\end{eqnarray}
and its inverse transformation
\begin{eqnarray}
\alpha_\nu = w_{\nu}^T\Psi ,\ \ \alpha_\nu^\dagger = w_{-\nu}^T\Psi.
\end{eqnarray}
The relation $\alpha_{-\nu}=\alpha^\dagger_\nu$ holds because of PHS.
Bogoliubov quasiparticles obey commutation relations of fermions so that $\{\alpha_\nu,\alpha_\mu^\dagger\} = \delta_{\nu,\mu}$ and $\{\alpha_\nu,\alpha_\mu\}=0$ are satisfied.
The mean-field Hamiltonian is diagonalized as
\begin{eqnarray}
\mathcal{H} = \frac{1}{2}\sum_{\nu>0} E_\nu(\alpha_\nu^\dagger\alpha_\nu-\alpha_\nu\alpha_\nu^\dagger).
\end{eqnarray}

In order to calculate the pair potential, we introduce an imaginary time Green's function, $\hat G(\tau) = -\langle T_\tau \Psi(\tau)\Psi^\dagger(0)\rangle$.
By the Bogoliubov transformation, the Green's function is expressed in the quasiparticle representation
\begin{eqnarray}
\hat G(\tau) = -\sum_{\nu} w_\nu w_\nu^T\left\langle T_\tau \alpha_\nu(\tau)\alpha_\nu^\dagger \right\rangle.
\end{eqnarray}

The pair potential defined in eq. (\ref{3.12}) is calculated from the anomalous part of the Green's function, i.e.,
\begin{eqnarray}
\Delta^{\tau\tau'}_{\sigma\sigma'}(i,j)
&=&\nonumber -g^{\tau\tau'}_{\sigma\sigma'}(i,j)G_{i\tau\sigma e,j\tau'\sigma'h}(\tau=+0)\\
&=&\nonumber
 g^{\tau\tau'}_{\sigma\sigma'}(i,j)\sum_{\nu>0}
\{
\left[1-f(E_\nu)\right]u_\nu(i\tau\sigma)v_\nu^*(j\tau'\sigma')\\
&+& f(E_\nu)v_\nu^*(i\tau\sigma)u_\nu(j\tau'\sigma') 
\}\label{order_parameter_determination},
\end{eqnarray} 
where \crd{the} two vectors $u$ and $v$ are introduced\crd{:}
\begin{eqnarray}
u_\nu(i\tau\sigma)=w_\nu(i\tau\sigma e), v_\nu(i\tau\sigma)=w_\nu(i\tau\sigma h).
\end{eqnarray}
Here, $u_\nu$ and $v_\nu$ correspond to the amplitudes of the particle and hole wave functions\crd{, respectively.}
In eq.(\ref{order_parameter_determination}), $f(E)=(e^{\beta E}+1)^{-1}$ is the Fermi-Dirac distribution function.
\crd{Note} that $u_\nu^*=v_{-\nu}$ is satisfied because of PHS.

\subsection{Symmetry of the \crd{o}rder \crd{p}arameter}
\rd{We first consider the impurity-free case and this subsection is devoted to remarks satisfied in the absence of impurities.}
We analyze the order parameter, which is determined from the real-space BdG equation \crd{eq.} \rd{(\ref{order_parameter_determination})}.
{In this subsection}, we concentrate on the order parameter $\hat{\tilde\Delta}$,
which does not contain the interaction coefficents $g^{\tau\tau'}_{\sigma\sigma'}(i,j)$ introduced in eq. (\ref{Ineraction})\crd{:}
\begin{eqnarray}
\tilde\Delta^{\tau\tau'}_{\sigma\sigma'}(i,j) \equiv \langle c_{i\tau\sigma}c_{j\tau'\sigma'} \rangle.
\end{eqnarray}
This order parameter $\hat{\tilde\Delta}$ represents the superfluid density.

The interaction parameter in eq. (\ref{Ineraction}) is 
\rd{\begin{eqnarray}
g^{EE}_{\uparrow\downarrow}(i,i)=g^{EE}_{\downarrow\uparrow}(i,i)=g^{HH}_{\uparrow\downarrow}(i,i)=g^{HH}_{\downarrow\uparrow}(i,i)=2\mathrm{\ eV}\crd{;}
\nonumber\\
\label{int_param}
\end{eqnarray}}
for the other components, $g^{\tau\tau'}_{\sigma\sigma'}(i,j)=0$.
By choosing the interaction coefficients as in eq. (\ref{int_param}),
we only consider onsite attractive interactions between
electrons in the same subband. 
We take the system size as $N_x=N_y=20$ \crd{and} $N_z=8$ in this subsection.

First, we show \crd{the} onsite components $\hat{\tilde\Delta}(i,i)$ of the order parameter obtained
\crd{using} the BdG equation.
For the onsite component, the result of the calculation is
\begin{eqnarray}
\hat{\tilde\Delta}(i,i)&=&\nonumber
\begin{pmatrix}
0 & 0 & \eta_1(z) & \mathrm{i}\eta_2(z)\\
0 & 0 & -\mathrm{i}\eta_2(z) & \eta_3(z)\\
-\eta_1(z) & \mathrm{i}\eta_2(z) & 0 & 0\\
-\mathrm{i}\eta_2(z) & -\eta_3(z) & 0 & 0\\
\end{pmatrix},\\
\eta_1(1)&=&1.45\times10^{-3},\eta_2(1)=1.35\times10^{-3}, \nonumber \\ 
\eta_3(1)&=&1.05\times10^{-3}.
\end{eqnarray}
where $\eta_1$ and $\eta_3$ are even functions of $z$\crd{,} while $\eta_2$ is an odd function of $z$ with the symmetry center \rd{at} $z=4.5$, i.e., the \rd{relations} $\eta_1(z) = \eta_1(9-z)$, $\eta_2(z-9) = -\eta_2(9-z)$, and $\eta_3(z-9) = \eta_3(9-z)$ hold.
The amplitude of each $\eta$
\crd{decreases} when $z$ is nearer to the the symmetry center, 
for example,
$\eta_1(1)= 1.45\times 10^{-3},
\eta_1(2)= 3.05\times 10^{-4},
\eta_1(3)= 2.39\times 10^{-5},
\eta_1(4)= 1.30\times 10^{-6}$\crd{;} therefore
$$\eta_1(1)>\eta_1(2)>\eta_1(3)>\eta_1(4),$$
holds.
This amplitude dependences on $z$ imply that the Cooper \rd{pairings} are mainly formed from the helical Dirac electrons, which are localized at the surface.
Since $\hat{\tilde\Delta}(i,i)$ is an antisymmetric matrix, each onsite order parameter $\eta$ is a singlet component.
Here, $\hat{\tilde\Delta}(i,i)$ is independent of $x$ and $y$ because of translational symmetry.

Then, we show a part of \crd{the} off-site components $\hat{\tilde\Delta}(i,i+e_x)$ of the order parameter obtained
\crd{using} the BdG equation.
Here, $\hat{\tilde\Delta}(i,i+e_x)$ represents the coherence between the two sites neighboring in \crd{the} $x$\crd{-}direction.
\rd{The} result of the calculation is 
\begin{eqnarray}
\hat{\tilde\Delta}(i,i+e_x)&=&\nonumber
\begin{pmatrix}
-\eta'_1(z) & \mathrm{i}\eta'_2(z) & \eta'_3(z) & \mathrm{i}\eta'_4(z)\\
\mathrm{i}\eta'_2(z) & \eta'_5(z) & -\mathrm{i}\eta'_4(z) & \eta'_6(z)\\
-\eta'_3(z) & \mathrm{i}\eta'_4(z) & -\eta'_1(z) & -\mathrm{i}\eta'_2(z)\\
-\mathrm{i}\eta'_4(z) &-\eta'_6(z)& -\mathrm{i}\eta'_2(z) & \eta'_5(z)\\
\end{pmatrix},\\
\eta'_1(1)&=&\nonumber 1.63\times10^{-4},\eta'_2(1)=1.71\times10^{-4},\\
\eta_3(1)&=&\nonumber 1.18\times10^{-3}, \eta'_4(1)=1.02\times10^{-3},\\
\eta'_5(1)&=& 1.60\times10^{-4},\eta_6(1)=1.12\times10^{-3},
\end{eqnarray}
where $\eta'_2$, $\eta'_3$\crd{,} and $\eta'_6$ are even functions of $z$\crd{,} while $\eta'_1$, $\eta'_4$\crd{,} and $\eta'_5$ are odd functions of $z$ with the symmetry center \rd{at} $z=4.5$.
Since $\eta'_1$, $\eta'_2$\crd{,} and $\eta'_5$  are symmetric components of the above matrix, they are triplet components.
\crd{In contrast,} $\eta'_3$, $\eta'_4$\crd{,} and $\eta'_6$ are antisymmetric components of the matrix and
are singlet components.

\crd{The above} results support \crd{the notion}
that an onsite $s$-wave interaction induces the order parameter with
\crd{a} mixture of singlet and triplet components, which agrees with the results of the idealistic helical Dirac electron model in \S 2.

Generally, the order parameter is written \rd{as}
\begin{eqnarray}
&&\hat{\tilde\Delta}(x,y,z;x+t_xe_x,y+t_y e_y,z)  \nonumber\\
&=& \nonumber
A(t,z)(t_xI\otimes I\rd{-}t_yI\otimes(\mathrm i\sigma_z))\\
&+&\nonumber
B(t,z)(t_x\tau_z\otimes I\rd{-}t_y\tau_z\otimes(\mathrm i\sigma_z))\\
&+&\nonumber
C(t,z)(t_x\tau_x\otimes (-\mathrm i\sigma_z)\rd{-}t_y\tau_x\otimes I)\\
&+&\nonumber
D(t,z)I\otimes (\mathrm i\sigma_y)+E(t,z)\tau_z\otimes(\mathrm i\sigma_y)\\
&+&F(t,z)\tau_y\otimes \sigma_x, \label{General_formation}
\end{eqnarray}
where $t=\sqrt{t_x^2+t_y^2}$. 
The above results are \crd{in} the case of $t_x=t_y=0$ and $t_x=1,t_y=0$.
Here, $C$, $D$\crd{,} and $E$ are even functions\crd{,}
while $A$, $B$\crd{,} and $F$ are odd functions of $z$.
In the above \crd{equation}, $\sigma$ acts
\crd{on} \rd{the} spin basis and the pseudospin Pauli matrix $\tau$ acts
\crd{on} the basis of $E$ and $H$ subbands.

According to the transformation rule in Appendix 1, the order parameter of the form in eq. (\ref{General_formation}) is invariant under a set of symmetry operations
\crd{that} belong to $\mathrm{D_{4h}}$.
Since the noninteracting original Hamiltonian is invariant under the symmetry operations in $\mathrm{D_{4h}}$,
the invariance of the order parameter under the operation
\crd{indicates} that the SC  realized in the model breaks no additional spatial symmetries.

\subsection{Stability against \crd{i}mpurities}
Now\crd{,} we analyze the impurity concentration
and
impurity strength dependence\crd{s} of the pair potential for the tight-binding model of $\mathrm{Bi_2Se_3}$.
We assume that impurities are located on one of the two surfaces.
We concentrate on
\crd{a} pair potential
\crd{on a} surface with impurities.
We define
\crd{a} surface with impurities as $z=1$.


By using the method
\crd{of} efficiently compensating \crd{for}
the change in the density of states (DOS) explained in \rd{\S A.3} in the case of the reference systems,
we analyze the impurity concentration dependence of the pair potential of $\mathrm{Bi_2Se_3}$.
Here\crd{, note}
that there are 16 components in the BdG equation for $\mathrm{Bi_2Se_3}$.
Therefore, one might speculate that the method
\crd{of} compensating \crd{for} the change in the DOS for the BdG equation
\crd{is} complicated.
However, by choosing the Hamiltonian parameters used in
\S 3.2, we can focus on $\Delta^{EE}_{\uparrow\downarrow}$ and $\Delta^{HH}_{\uparrow\downarrow}$.
This is because we only consider onsite attractive interactions between
electrons in the same subband.
In the following discussion, we concentrate on \crd{the} \rd{dependences of $\Delta^{EE}_{\uparrow\downarrow}$ on}
the scattering strength and impurity concentration, \rd{and $\Delta^{EE}_{\uparrow\downarrow}$ is written as $\Delta$}.

\rd{In order to describe the behavior of the pair potential purely due to
relaxation processes, we concentrate on the value
\begin{eqnarray}
1+\frac{\delta\Delta(u,n_\mathrm{i})}{\Delta_0}= \frac{\langle\langle\Delta(u,n_\mathrm{i})\rangle_x\rangle_\mathrm{imp}}{\langle\Delta_\mathrm{(DOS)}(u,n_\mathrm{i})\rangle_\mathrm{imp}} \label{3.27},
\end{eqnarray}
which represents the relative value of the pair potential due to the relaxation.
In eq. (\ref{3.27}), the quantities $\Delta_0$, $\delta\Delta$, $\Delta(u,n_\mathrm{i})$, and $\Delta_{(\mathrm{DOS})}(u,n_\mathrm{i})$ represent
\crd{the} impurity-free pair potential, \crd{the} change
\crd{in} pair potential purely due to the relaxation, the result of the BdG calculation, and the relaxation-ignored pair potential, respectively.
The definitions of these quantities are introduced in \rd{\S A.3}.
\crd{Note} that $\Delta(u,n_\mathrm{i})$ depends on the site and
impurity configuration\crd{,} and $\Delta_{(\mathrm{DOS})}(u,n_\mathrm{i})$ depends on the \bl{impurity configuration}.
Moreover, $\langle\cdots\rangle_x$ and $\langle\cdots \rangle_\mathrm{imp}$ represent the average over sites and impurity configuration, respectively, whose definitions are
\crd{given} in \rd{\S A.3.}}

\rd{In the model, we estimate $\Delta_{\mathrm{(DOS)}}$ as
\begin{eqnarray}
&&\Delta_\mathrm{(DOS)}(u,n_\mathrm{i}) =\nonumber \Delta_0 \\
&\times&\frac{ \left[ \sum_{\nu}\frac{\Delta_\mathrm{(DOS)}(u,n_i)}{\sqrt{E_\nu(u,n_i)^2+\Delta_\mathrm{(DOS)}(u,n_i)^2}}\rho(E_\nu(u,n_i),z=1)\right]}
{\left[\sum_{\nu}\frac{\Delta_0}{\sqrt{E_{0\nu}^2+\Delta_0^2}}\rho^{(0)}(E_{0\nu},z=1)\right]},\nonumber\\
\end{eqnarray}
where $\rho^{(0)}(E,z=1)$ is the DOS at the surface without impurities\crd{,} while $\rho(E,z=1)$ is that with impurities.
Here, $E_{0\nu}$ and $E_{\nu}(u,n_\mathrm{i})$ represent \crd{the} energy
\crd{spectra} without and with impurities\crd{, respectively}.
We note that $E_{\nu}(u,n_\mathrm{i})$ depends on the impurity configuration.}

We show the results of $1+\delta\Delta(u,n_\mathrm{i})/\Delta_0$ in Figs. 3 and 4.
Figure 3 shows the impurity concentration dependence of the pair potential for \rd{the TRS impurities}.
In the figure, the results for the scattering strength of the choices $u=0.1$ and $u=20$ are plotted.

For $u=0.1$, the inset in Fig. 3 shows that the reduction
\crd{in} the pair potential is proportional to the square of the impurity concentration, i.e., \rd{$\propto n_\mathrm{i}^2$} for small $n_\mathrm{i}$.
Since the order of the pair potential is $10^{-3}$, $u=0.1$ is much larger than the pair potential.
Therefore, it is \rd{beyond the applicability of} the AG theory.
The robustness of the SC obtained in the BdG calculation \rd{indicates the stability of the SC beyond the perturbative range}.
For $u=0.1$, the pair potential \rd{vanishes} at $n_\mathrm{i}\simeq 5\%$.
On the other hand, when the strength of impurity potential is $u=20$,
which
\crd{indicates} a strong impurity potential regarded \rd{practically} as a lattice defect, the pair potential vanishes at $n_\mathrm{i}\simeq 4\%$.

In Fig. 4, we compare the results for
TRS scattering
with
magnetic scattering\crd{.}
In the case of
magnetic scattering\crd{,}
the pair potential decreases linearly for small concentration\crd{s}.
We find that the reduction rate of the pair potential \rd{depends on the direction along $x$ or $z$ of the magnetization
\crd{for} magnetic impurities.}
We deduce that
\crd{a} difference
\crd{between} the two \rd{directions} remains \rd{even} at
\crd{an} impurity concentration
\crd{range higher} than 0.25\% in spite of the large error bars.
\rd{In fact,} the difference \rd{in} the impurity concentration dependences at 0.25\% \bl{ is reliable and statistically meaningful.}
Actually, according to the AG theory, the reduction rate\crd{s} of the pair potential
\crd{are} the same in
both cases.
Therefore,
\crd{a} difference \rd{in} the impurity concentration dependence
arises from
higher\crd{-}order perturbations.
This difference is mainly caused by the difference \rd{in} the change in DOS\crd{,} which depends on the polarization.\cite{cit:PRB81_233405}

\begin{figure}[htbp]
\begin{center}
\includegraphics[width=80mm]{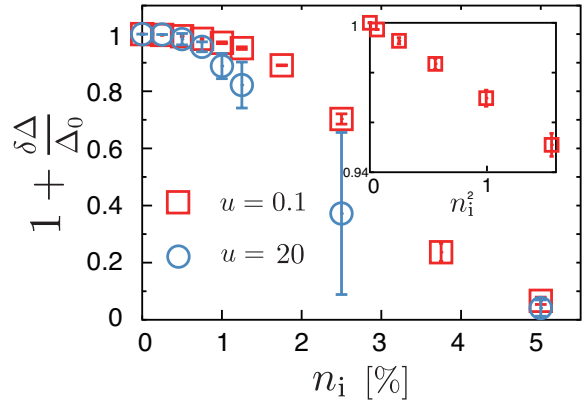}
\caption{(Color online)
Impurity concentration dependence of pair potential for TRS scattering\crd{.}
The abscissa represents the impurity concentration $n_\mathrm{i}$, while the ordinate represents $1+\delta\Delta(u,n_\mathrm{i})/\Delta_0$.
In the calculation, we fix  the system size at $N_xN_y=400$.
The squares and circles illustrate the results for
scattering\crd{s of} strength $u=0.1$ and $u=20$\crd{, respectively}.
In the inset, the same quantity $1+\delta\Delta(u,n_\mathrm{i})/\Delta_0$ is plotted as a function of the square of the impurity concentration $n_\mathrm{i}^2$
for $u=0.1$.}
\label{default}
\end{center}
\end{figure}

\begin{figure}[htbp]
\begin{center}
\includegraphics[width=80mm]{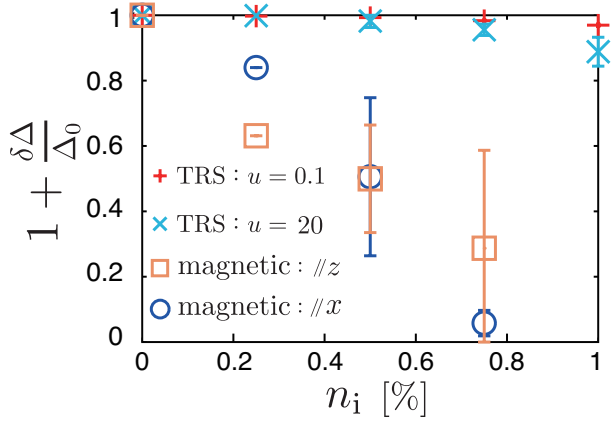}
\caption{(Color online)
Impurity concentration dependence of pair potential for TRS and magnetic impurity scatterings. 
The abscissa represents the impurity concentration $n_\mathrm{i}$, while the ordinate represents $1+\delta\Delta(u,n_\mathrm{i})/\Delta_0$.
In the calculation, we fix the system size at $N_xN_y=400$.
The plus and cross illustrate the results for the TRS impurity scattering
when $u=0.1$ and $u=20$\crd{, respectively}.
The squares and circles illustrate the results for the magnetic impurity scattering\crd{,}
which \crd{were} polarized along the $x$\crd{-} and $z$\crd{-}directions.}
\label{default}
\end{center}
\end{figure}

\subsection{Spatial \crd{s}tructure}
In this section, we analyze spatial structures \rd{induced} by an impurity.
We focus on the spatial structures of the pair potential and wave functions of Bogoliubov quasiparticles.
To study the pair potential, we obtain a configuration of induced currents around an impurity. 
To study wave functions of
quasiparticles, we analyze the properties of bound states around an impurity.
\rd{Here, we assume that}
the impurity is located at $(x,y)=(0,0)$ and
that the impurity potential $u$ is strong.

In the calculation, we choose the parameters of $\mathcal{H}$ as
\rd{\begin{eqnarray}
&&A_x=1\mathrm{\ eV},A_z=0.5\mathrm{\ eV},B_x=1.5\mathrm{\ eV},B_z=0.3\mathrm{\ eV},\nonumber\\
&&D_x=D_z=0,M=0.5\mathrm{\ eV} ,\\
&&g^{\tau\tau'}_{\sigma\sigma'}(\bm r,\bm r)=2\mathrm{\ eV}\ \  (\mathrm{onsite\ components}),\\
&& g^{\tau\tau'}_{\sigma\sigma'}(\bm r,\bm r')=1\mathrm{\ eV} \ \ (\bm r\ \mathrm{and\ }\bm r'\ \mathrm{are\ the\ nearest\ neighbors}).\nonumber\\
\end{eqnarray}}
We take the system size \crd{as} $N_x=N_y=20$ and $N_z=8$.

First, we show the spatial dependence of the onsite $s$-wave components around the impurity.
\rd{Here, $s$-wave components are defined as
\begin{eqnarray}
\Delta^{\tau\tau'}_{\uparrow\downarrow s}(\bm r) &\equiv& \Delta^{\tau\tau'}_{\uparrow\downarrow}(\bm r,\bm r)\nonumber \\
\Delta^{\tau\tau'}_{\downarrow\uparrow s}(\bm r) &\equiv& \Delta^{\tau\tau'}_{\downarrow\uparrow}(\bm r,\bm r).
\end{eqnarray}}
Figure 5
\crd{shows} the spatial dependence of one of those components \rd{$\Delta^{EE}_{\uparrow\downarrow s}(\bm r)$}.
The amplitude of \rd{$\Delta^{EE}_{\uparrow\downarrow s}(\bm r)$} nearly vanishes at the impurity site.
\bl{This is because the impurity potential is so strong that it is \crd{nearly} regarded as a lattice defect and the electron density is extremely small at the impurity site.}

\begin{figure}[htbp]
\begin{center}
\includegraphics[width=80mm]{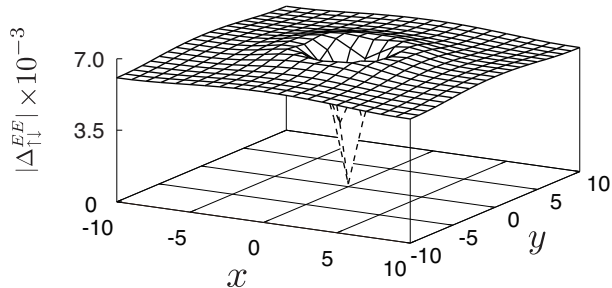}
\caption{Spatial dependence of $|\Delta^{EE}_{\uparrow\downarrow}|$.}
\label{default}
\end{center}
\end{figure}

We then show the spatial dependence of the onsite $p$-wave components.
Figure 6 shows \rd{$\Delta^{EH}_{\uparrow\uparrow p}(\bm r)\equiv\Delta^{EH}_{\uparrow\uparrow }(\bm r,\bm r)$} which is one of the onsite $p$-wave components.
This component reflects the existence of inter-orbital Cooper pairings.
At the impurity site, this value is zero in agreement with the symmetry of the original Hamiltonian with an impurity potential.
Because the system has the $C_4$ rotational symmetry around the impurity site, the impurity site is a singular point of the vector field, \rd{$(\mathrm{Re}[\Delta^{EH}_{\uparrow\uparrow p}(\bm r)],\mathrm{Im}[\Delta^{EH}_{\uparrow\uparrow p}(\bm r)])$}, where \rd{$\Delta^{EH}_{\uparrow\uparrow p}(\bm r)$} is zero.
Without the impurity, this component \rd{vanishes} because the lattice translational symmetries along the $x$\crd{-} and $y$\crd{-}directions exist and every site is the rotational center.

Figure 7 shows the spatial configurations of two normalized vector fields,
\rd{$\bm F^{EH}_{\uparrow\uparrow}(\bm r)/|\bm F^{EH}_{\uparrow\uparrow}(\bm r)|$ and $\bm F^{EH}_{\downarrow\downarrow}(\bm r)/|\bm F^{EH}_{\downarrow\downarrow}(\bm r)|$}\crd{,} where
\begin{eqnarray}
\rd{\bm F^{EH}_{\uparrow\uparrow}(\bm r) = \left(\mathrm{Re}[\Delta^{EH}_{\uparrow\uparrow p}(\bm r)],\mathrm{Im}[\Delta^{EH}_{\uparrow\uparrow p}(\bm r)]\right)} ,
\end{eqnarray}
and
\begin{eqnarray}
\rd{\bm F^{EH}_{\downarrow\downarrow}(\bm r)=\left(\mathrm{Re}[\Delta^{EH}_{\downarrow\downarrow p}(\bm r)],\mathrm{Im}[\Delta^{EH}_{\downarrow\downarrow p}(\bm r)]\right)} .
\end{eqnarray}
Since we fix the length of the arrows representing the vector fields, only the directions of the arrows are meaningful, which correspond to the complex phases of \rd{$\Delta^{EH}_{\uparrow\uparrow p}(\bm r)$ or $\Delta^{EH}_{\downarrow\downarrow p}(\bm r)$}.
At
singular points, where $|\bm F|=\vec 0$ holds, the arrows are not plotted.
In each component, there exist vortices at $(0,0)$, $(0,10)$, $(10,0)$\crd{,} and $(10,10)$.
In the $\Delta^{EH}_{\uparrow\uparrow}$ component, the vortices at $(0,0)$, $(0,10)$\crd{,} and $(10,0)$ \rd{are} circulating clockwise while the vortex at $(10,10)$ is circulating counterclockwise.
\crd{In contrast,} the vortices of $\Delta^{EH}_{\downarrow\downarrow}$ have
opposite chiralities, \rd{i.e., 
\begin{eqnarray}
&&\frac{1}{2\pi}\oint d\bm s\cdot\bm F_{\uparrow\uparrow}^{EH}(\bm r)/|\bm F_{\uparrow\uparrow}^{EH}(\bm r)| \nonumber\\
&=& 
-\frac{1}{2\pi}\oint d\bm s\cdot\bm F_{\downarrow\downarrow}^{EH}(\bm r)/|\bm F_{\downarrow\downarrow}^{EH}(\bm r)|. \label{3.33}
\end{eqnarray}
In eq. (\ref{3.33}), the chiralit\crd{ies} of $\Delta^{EH}_{\uparrow\uparrow p}(\bm r)$ and $\Delta^{EH}_{\downarrow\downarrow p}(\bm r)$ around a vortex
\crd{are respectively} defined as $\frac{1}{2\pi}\oint d\bm s\cdot\bm F_{\uparrow\uparrow}^{EH}(\bm r)/|\bm F_{\uparrow\uparrow}^{EH}(\bm r)|$ and $\frac{1}{2\pi}\oint d\bm s\cdot\bm F_{\downarrow\downarrow}^{EH}(\bm r)/|\bm F_{\downarrow\downarrow}^{EH}(\bm r)|$ with an integral path around the vortex.
}

\begin{figure}[htbp]
\begin{center}
\includegraphics[width=80mm]{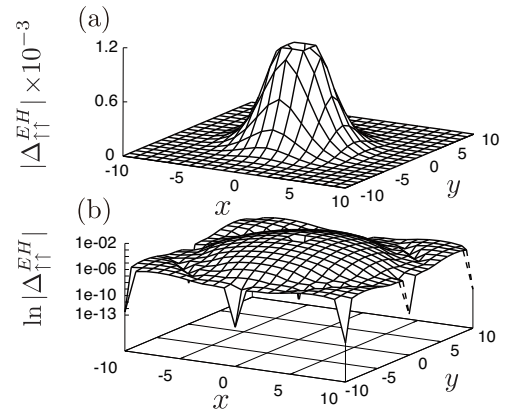}
{\caption{Spatial dependences of (a) $|\Delta^{EH}_{\uparrow\uparrow}|$ and (b) $\ln|\Delta^{EH}_{\uparrow\uparrow}|$.
Here, $|\Delta^{EH}_{\uparrow\uparrow}|=0$ holds at the origin and is not plotted in the lower panel.}} 
\label{default}
\end{center}
\end{figure}

\begin{figure}[htbp]
  \begin{center}
   \includegraphics[width=80mm]{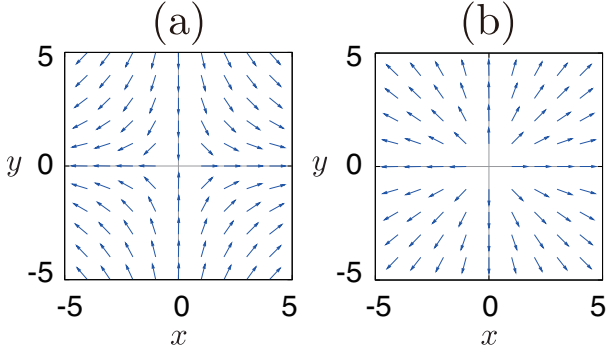}
  \end{center}
\caption{{(Color online) Spatial configuration of normalized vector fields, (a) $\bm F^{EH}_{\uparrow\uparrow}/|\bm F^{EH}_{\uparrow\uparrow}|$ and (b) $\bm F^{EH}_{\downarrow\downarrow}/|\bm F^{EH}_{\downarrow\downarrow}|$.
The direction of the arrow represents the complex phase of $\Delta^{EH}_{\uparrow\uparrow}$ or $\Delta^{EH}_{\downarrow\downarrow}$.
At
singular points, where $|\bm F|=\vec 0$ holds, the arrows are not plotted.} }
\end{figure}

\newpage
In the present model, the off-site $p$-wave components \rd{$\Delta_{p_x}$ and $\Delta_{p_y}$} exist simultaneously.
\rd{They} are calculated from the off-site pair potentials as 
\begin{eqnarray}
\Delta_{\sigma\sigma' p_i}^{\tau\tau'}(\bm r) &=& \frac{1}{2}\left[\Delta^{\tau\tau'}_{\sigma\sigma'}(\bm r,\bm r+\bm e_i)-\Delta^{\tau\tau'}_{\sigma\sigma'}(\bm r,\bm r-\bm e_i)\right] \label{pcomponent1}\ \ (i=x,y). \nonumber \\
\end{eqnarray}

Then we introduce the representation of $p_x\pm\mathrm{i} p_y$ components as
\rd{\begin{eqnarray}
\eta^{\tau\tau'}_{\sigma\sigma'\pm}(\bm r) = \frac{1}{2}(\Delta_{\sigma\sigma' p_x}^{\tau\tau'}(\bm r) \mp\mathrm{i}\Delta_{\sigma\sigma' p_y}^{\tau\tau'}(\bm r)),
\end{eqnarray}}
where $\eta_\pm$ corresponds to the $p_x\pm\mathrm{i}p_y$ component.
This representation is useful when describing the pair potential of chiral $p$-wave-like SCs.

\dark{Without impurities, $(\Delta^{\tau\tau'}_{\uparrow\uparrow p_x},\ \Delta^{\tau\tau'}_{\uparrow\uparrow p_y})$ consists of
$\eta_-$ components and  $(\Delta^{\tau\tau'}_{\downarrow\downarrow p_x},\ \Delta^{\tau\tau'}_{\downarrow\downarrow p_y})$ consists of
$\eta_+$ components.
These results are consistent with the results in
\crd{\S} 3.1.
For example, since $\Delta^{EE}_{\uparrow\uparrow p_x}(x,y,z=1)=A(1,1)+B(1,1)$ and $\Delta^{EE}_{\uparrow\uparrow p_y}(x,y,z=1)=-\mathrm{i}[A(1,1)+B(1,1)]$ are derived from eq. (\ref{General_formation}) without impurities, $\eta_{\uparrow\uparrow +}^{EE}$ vanishes and $\eta_{\uparrow\uparrow -}^{EE}$ remains.}

We show the spatial structures of $\Delta^{EE}_{\uparrow\uparrow}$ and $\Delta^{EE}_{\downarrow\downarrow}$.
Figure 8  shows $\eta^{EE}_{\uparrow\uparrow-}$, which is the dominant component of $\Delta^{EE}_{\uparrow\uparrow}$.
This configuration is the same as that of $\eta^{EE}_{\downarrow\downarrow+}$.
Figure 9 shows \crd{the} absolute values of $\eta^{EE}_{\uparrow\uparrow+}$
induced by the impurity.
From the spatial configuration
\crd{on} the logarithmic scale, we find that there are six vortices in the system.
Figure 10 shows the spatial configurations of two normalized vector fields, $\bm F^{EE}_{\uparrow\uparrow+}/|\bm F^{EE}_{\uparrow\uparrow+}|$ and $\bm F^{EE}_{\downarrow\downarrow-}/|\bm F^{EE}_{\downarrow\downarrow-}|$, where
\rd{\begin{eqnarray}
\bm F^{EE}_{\uparrow\uparrow+}(\bm r) &=& \left(\mathrm{Re}[\eta^{EE}_{\uparrow\uparrow+}(\bm r)],\mathrm{Im}[\eta^{EE}_{\uparrow\uparrow+}(\bm r)]\right),
\end{eqnarray}}
and
\rd{\begin{eqnarray}
\bm F^{EE}_{\downarrow\downarrow-}(\bm r)=\left(\mathrm{Re}[\eta^{EE}_{\downarrow\downarrow-}(\bm r)],\mathrm{Im}[\eta^{EE}_{\downarrow\downarrow-}(\bm r)]\right).
\end{eqnarray}}
Since we fix the length of the arrows representing the vector fields, only the directions of the arrows are meaningful, which correspond to the complex phases of \rd{$\eta^{EE}_{\uparrow\uparrow+}(\bm r)$ or $\eta^{EE}_{\downarrow\downarrow-}(\bm r)$}.
At singular points, the arrows are not plotted as
\crd{in} Fig. 7.
Since \rd{$\eta^{EE}_{\uparrow\uparrow +}$ and $\eta^{EE}_{\downarrow\downarrow -}$} have
opposite \rd{chiralities, i.e., 
\begin{eqnarray}
&&\frac{1}{2\pi}\oint d\bm s\cdot\bm F_{\uparrow\uparrow +}^{EE}(\bm r)/|\bm F_{\uparrow\uparrow +}^{EE}(\bm r)| \nonumber\\
&=& 
-\frac{1}{2\pi}\oint d\bm s\cdot\bm F_{\downarrow\downarrow -}^{EE}(\bm r)/|\bm F_{\downarrow\downarrow -}^{EE}(\bm r)|, \label{3.38}
\end{eqnarray}}
\crd{the} induced circular electric current
of each spin component flows in
opposite directions.
\rd{In eq.(\ref{3.38}), the chiralit\crd{ies} of $\eta^{EE}_{\uparrow\uparrow +}$ and $\eta^{EE}_{\downarrow\downarrow -}$ around a vortex
\crd{are respectively} \bl{defined} as $\frac{1}{2\pi}\oint d\bm s\cdot\bm F_{\uparrow\uparrow +}^{EE}(\bm r)/|\bm F_{\uparrow\uparrow +}^{EE}(\bm r)|$
and $\frac{1}{2\pi}\oint d\bm s\cdot\bm F_{\downarrow\downarrow -}^{EE}(\bm r)/|\bm F_{\downarrow\downarrow -}^{EE}(\bm r)|$\crd{,} with an integral path around the vortex.}
These circulating currents represent spin currents.

\begin{figure}[htbp]
\begin{center}
\includegraphics[width=80mm]{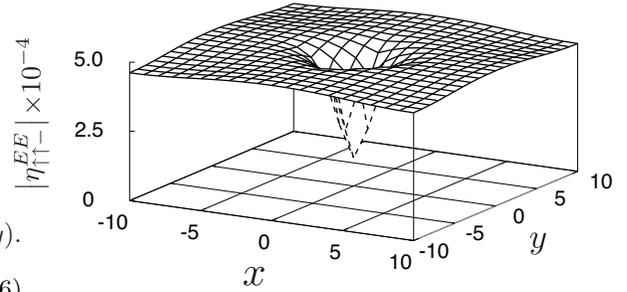}
\caption{{Spatial dependence of $|\eta^{EE}_{\uparrow\uparrow-}|$}, which is the dominant component of $\Delta^{EE}_{\uparrow\uparrow}$.
This configuration is the same as $|\eta^{EE}_{\downarrow\downarrow+}|$.}
\label{default}
\end{center}
\end{figure}

\begin{figure}[htbp]
  \begin{center}
   \includegraphics[width=80mm]{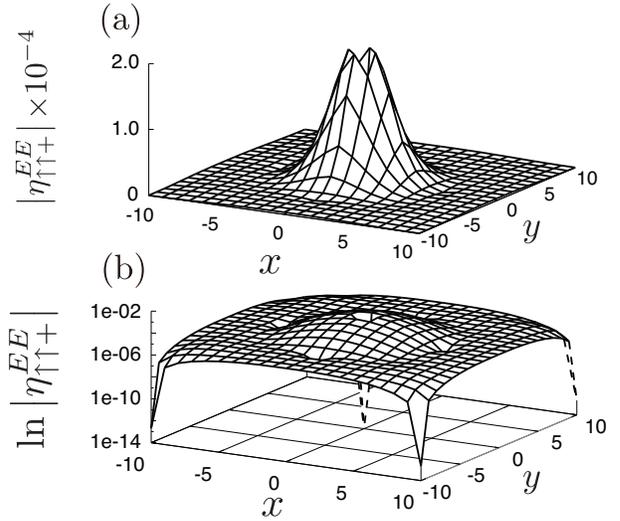}
  \end{center}
 \caption{{Spatial dependence\crd{s} of} (a) $|\eta^{EE}_{\uparrow\uparrow+}|$ and (b) $\ln|\eta^{EE}_{\uparrow\uparrow+}|$.
{These configurations are the same as $|\eta^{EE}_{\downarrow\downarrow-}|$ and $\ln|\eta^{EE}_{\downarrow\downarrow-}|$ respectively}. }
\end{figure}

\begin{figure}[htbp]
  \begin{center}
   \includegraphics[width=80mm]{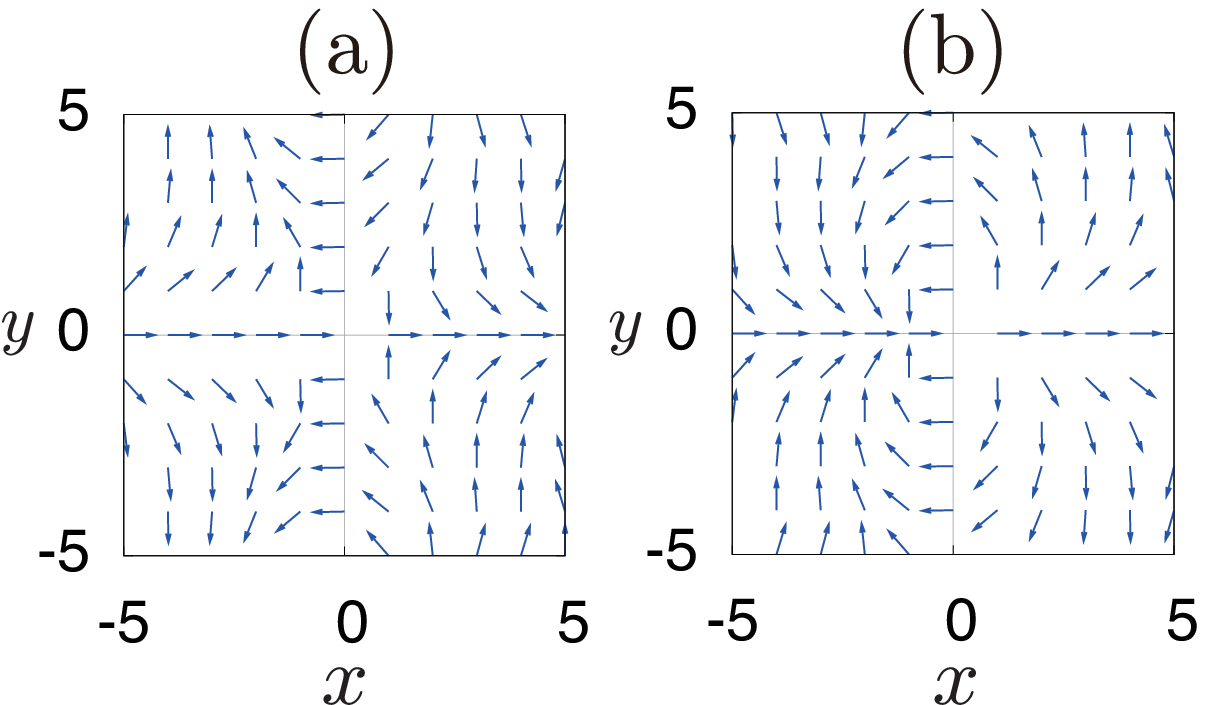}
  \end{center}
{\caption{(Color online) Spatial configurations of normalized vector fields, (a) $\bm F^{EE}_{\uparrow\uparrow+}/|\bm F^{EE}_{\uparrow\uparrow+}|$ and (b) $\bm F^{EE}_{\downarrow\downarrow-}/|\bm F^{EE}_{\downarrow\downarrow-}|$. 
The direction of the arrow represents the complex phase of $\eta^{EE}_{\uparrow\uparrow+}$ or $\eta^{EE}_{\downarrow\downarrow-}$.
At
singular points, where $|\bm F|=\vec 0$ holds, the arrows are not plotted.
The two opposite components have
opposite chiralities.} }
\end{figure}

\rd{In this subsection, we observe \crd{the} spatial structures of typical pair-potential components.
For the complex phase of $s$-wave and $p$-wave components, qualitatively different spatial configurations are obtained, both of which are allowed in the original symmetry of the system.
\crd{In particular,}
the complex phase of $p$-wave components around impurities should be observed as spin currents around them.
}

\subsection{Wave \crd{f}unction of Bogoliubov \crd{q}uasi-\crd{p}article}
Now\crd{,} we study the energy spectra and wave functions of a Bogoliubov quasiparticle composed of SSs.
In the present calculation, there are eight states near the Fermi energy. 
Without impurities these eight states are degenerate since the system has
\crd{a} four-fold rotational symmetry and
\crd{an} inversion symmetry.
By introducing an impurity potential, these states split.
In Fig. 11, we
\crd{show} the ratio of \rd{the energy level splitting $E_{\mathrm{imp}}$ to the BCS energy gap $E_{\mathrm{BCS}}$} caused by an impurity when the system size
\crd{changes.}
Since $E_\mathrm{imp}/E_{\mathrm{BCS}}$ is scaled to zero in the thermodynamic limit, it supports \crd{the notion} that the impurity level does not split off from the gap edge.
This means that the state does not \rd{develop into} a mid-gap bound state.

\begin{figure}[htbp]
\begin{center}
\includegraphics[width=70mm]{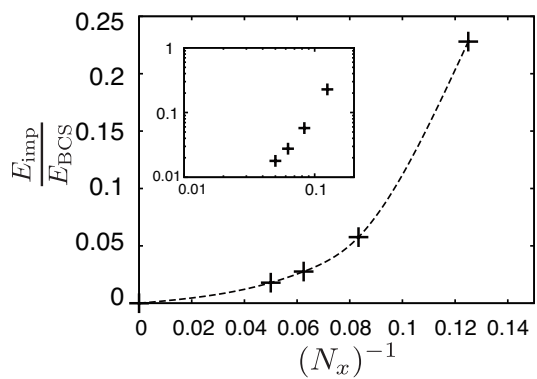}
\caption{Relation\crd{ship} between system size and ratio of \rd{energy level splitting $E_{\mathrm{imp}}$ to BCS energy gap $E_{\mathrm{BCS}}$} caused by an impurity.
\textcolor{black}{Inset shows
\crd{a} log-log plot for the same data.}
}
\label{default}
\end{center}
\end{figure}

\begin{figure}[htbp]
\begin{center}
\includegraphics[width=70mm]{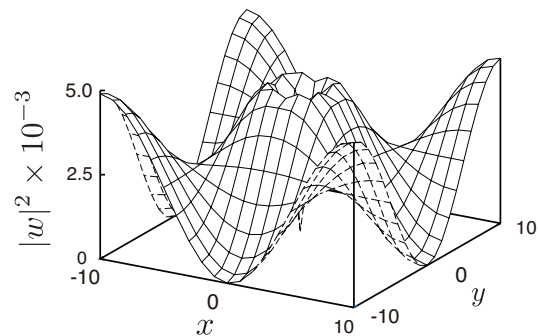}
\caption{{Amplitude of} the quasiparticle wave function {for} the state {whose energy is} the nearest to the Fermi energy.}
\label{default}
\end{center}
\end{figure}
In Fig. 12, the amplitude of the quasiparticle wave function
\crd{that} has
\crd{an} energy nearest to the Fermi energy is shown.
This wave function is spatially extended and supports
\crd{the notion} that\crd{,} in the thermodynamic limit, the state in Fig. 11 does not seem to \bl{reduce} to a mid-gap bound state\crd{,}
in agreement with the absence of
\crd{a} bound state.

If there \rd{were} an energy level of a mid-gap or gapless state in a SC,
it \crd{would}
not
\crd{be} able to penetrate
\crd{the} bulk due to \rd{the} existence of a SC gap.
Therefore, such a mid-gap state has to be bounded near the surfaces or edges of the SC.
The mid-gap state is referred to as an Andreev bound state (ABS).
\rd{Because the SC is a full gap SC in the \rd{2D surface}, the stability of
ABSs is related to the topology of the \bl{superconducting} gap;
\crd{i}f the \rd{superconducting} gap is topologically nontrivial,
ABSs \rd{should be} topologically protected and stable, and such a SC is called a topological SC\crd{, which}
is
\crd{analogous in stability of} SSs in TIs.}

Our result supports \crd{the notion} that there are no ABSs
\crd{that} are mid-gap states formed around the impurity.
The absence of ABSs represents \crd{the notion} that \crd{the} superconductivity gap induced by an $s$-wave attractive interaction on \crd{the} surfaces of TIs is topologically trivial,
i.e., the present SC is not a topological SC.
On the other hand, spinless chiral $p$-wave SCs, which have similar forms of pair potential to 
the present SCs, are topological SCs\cite{cit:PRB61_10267}.
This means that
SCs on \crd{the} surfaces of TIs and spinless chiral $p$-wave SCs are topologically different.

\section{Conclusion and Discussion}
In \S 2, we have shown that unconventional SCs induced by the $s$-wave attractive interaction on \crd{the} surfaces of TIs are robust
\crd{against} TRS disorders in idealistic Dirac electron models.
This is because of the cancellation of two phase factors, i.e., one from the pairing potential and the other from the scattering factor of Dirac electron\crd{s}.
\crd{In contrast,} unconventional SCs studied in the literature, such as
$d$-wave and chiral $p$-wave SC\crd{s}\cite{cit:RMP78_373}, are sensitively suppressed through scattering by a tiny concentration of impurities because of the phase factor of the pairing potential.

In \S 3, by numerically analyzing the Bogoliubov de-Gennes equation beyond the perturbative regime for
impurities, we have obtained the result that SCs \crd{induced} by the $s$-wave attractive interaction on \crd{the}
surfaces of TIs are stable against TRS impurities since the pair potential does not decrease linearly within
\crd{a} range of small concentration. 
This result is consistent with that
\crd{discussed in} \S 2.
Moreover, we have found that the robustness is observed even \rd{when} the impurity potential is \rd{strong} beyond the perturbation theory.
\rd{We} have \rd{also} found that the reduction rate
depend\crd{s} on the \rd{spin} polarization of magnetic impurities.
This is because \rd{the change} in the DOS depend\crd{s} on the polarization\cite{cit:PRB81_233405}.
This implies that the \rd{superconducting} gap in the present SC is topologically trivial, i.e., the present SC is a trivial SC.
Generally, \crd{the} existence of ABSs and \crd{the} stability of SCs are closely related, i.e., SCs
\crd{that} are stable against TRS impurities have ABSs around impurities.
According to our results, this relation \rd{appears to be satisfied in} the present SC on \crd{the} surfaces of TIs.

Finally, we describe issues left for future \crd{study}. 
In this \rd{paper}, we have studied the superconductivity stabilized by an internal $s$-wave interaction in order to clarify fundamental impurity effects.
Recently, however, superconductivity introduced by the proximity effect has also actively been studied\cite{cit:PRL100_096407,cit:arXiv1112.1772}.
\rd{Impurity} effects on the SC introduced by the proximity effect are also intriguing.

Moreover, stability against vortices is important.
This is because, in realizing a Majorana bound state around a vortex\cite{cit:PRL100_096407}, \crd{the} backscattering
of Dirac electrons due to a magnetic field, which breaks TRS, can be destructive \rd{in terms of} the stability of
SSs. 

\acknowledgement{
\crd{Numerical calculation was partly carried out at the Supercomputer Center, Institute for Solid State Physics,
The University of Tokyo.}
This work has been supported by Grant-in-Aid for Scientific
Research from MEXT Japan under the grant numbers
22104010 and 22340090. This work has also been financially
supported by MEXT HPCI Strategic Programs for
Innovative Research (SPIRE) and Computational Materials
Science Initiative (CMSI).
}

\sectiona
\appendix
\subsection{Symmetry \crd{o}peration}
Here, we discuss symmetry operations \textcolor{black}{that} preserve spatial symmetries that the Hamiltonian in eq. (\ref{Bi2Se3_Model}) has.
\textcolor{black}{\crd{The s}ymmetries of superconducting pair potentials are determined by irreducible representations of a group consisting of these symmetry operators.}

\textcolor{black}{First, we introduce \crd{a} matrix representation of symmetry operators}
as \textcolor{black}{follows:}
\textcolor{black}{A symmetry operator on a Hilbert space
is denoted as $\mathcal P$, while its matrix representation acting on creation (annihilation) operator-vectors
$\hat P$ ($\hat P^\dagger$) is defined as}
\begin{eqnarray}
\mathcal Pc^\dagger \mathcal P^{-1} &=&c^\dagger \hat P\textcolor{black}{,}\\
\mathcal Pc\mathcal P^{-1} &=& \hat P^{\dagger}c,
\end{eqnarray}
where $\hat P$ depends on \crd{the} choice of the basis.

Then \crd{the} BCS mean-field Hamiltonian is transformed \textcolor{black}{under a symmetry} operation $\mathcal P$ as
\begin{eqnarray}
\mathcal P \mathcal H\mathcal P^{-1} = 
\frac12\begin{pmatrix}c^\dagger \hat P & c^T \hat P \end{pmatrix}
\begin{pmatrix}
\hat H_0 & \hat\Delta\\
\hat\Delta^\dagger  & -\hat H_0^T
\end{pmatrix}
\begin{pmatrix}\hat P^\dagger c\\ \hat P^T(c^\dagger)^T  \end{pmatrix}.\label{transformation_rule}
\end{eqnarray}
If $\mathcal{PHP}^{-1}=\mathcal{H}$
, i.e., $\hat P \hat H_0 \hat P^\dagger = \hat H_0$ and $\hat P \hat \Delta \hat P^T = \hat \Delta$, 
is satisfied, the noninteracting Hamiltonian \rd{${\hat H}_{0}$}
and the order parameter \rd{matrix $\hat \Delta$} are both invariant under the operation $\mathcal{P}$.

\rd{The effective tight-binding Hamiltonian $\mathcal H_0$ of Bi$_2$Se$_3$ in \S 3}
is invariant under the symmetry operations belonging to \crd{the} point group $\mathrm{D_{4h}}$.
\rd{These symmetry operations are characterized by the}
character \bl{table given} in Table A.1.

With the choice of the basis employed in this paper, we obtain matrix representation \rd{for each symmetry operation in $\mathrm{D_{4h}}$,}
\rd{as shown} in Table\crd{s} A.2. and
A.3.
Table A.2 corresponds to the matrix acting on \crd{the} spin \textcolor{black}{indices}
and subbands \textcolor{black}{indices}
$E$ and $H$, and
Table A.3 corresponds to the matrix acting on spatial \textcolor{black}{coordinates.}
In Table A.2, \textcolor{black}{we note that}
$\hat\sigma$ acts on spin
\textcolor{black}{indices}
and $\hat\tau$ acts on subbands
\textcolor{black}{indices.}
\textcolor{black}{When we assume an $s$-wave attractive interaction in the effective model,
we obtain pair potentials belonging to the irreducible representation $\mathrm{A}_{1\mathrm {g}}$.}
\onecolumn
\begin{table}[htbp]
\caption{
Character table for $\mathrm{D_{4h}}$ point group.}
\begin{center}
\begin{tabular}{c|ccccccccccc}
\hline 
$\mathrm D_{4\mathrm h}$ 
& $E$ & $2C_4$ & $C_4^2$ & $2C_2'$ & $2C_2''$ & $I$ & $S_4$ & $\sigma_h$ & $2\sigma_v$ & $2\sigma_d$  & \rd{Example of basis}\\
\hline
$\mathrm{A}_{1\mathrm {g}}$ & $1$ & $1$  & $1$ & $1$  & $1$  & $1$ & $1$  & $1$ & $1$  & $1$ & $z^2\ \mathrm{or}\ x^2+y^2$ \\
$\mathrm{A}_{2\mathrm {g}}$ & $1$ & $1$  & $1$ & $-1$ & $-1$ & $1$ & $1$  & $1$ & $-1$ & $-1$ & $xy(x^2-y^2)$ \\
$\mathrm{B}_{1\mathrm {g}}$ & $1$ & $-1$ & $1$ & $1$  & $-1$ & $1$ & $-1$ & $1$ & $1$  & $-1$ & $x^2-y^2$\\
$\mathrm{B}_{2\mathrm {g}}$ & $1$ & $-1$ & $1$ & $-1$ & $1$  & $1$ & $-1$ & $1$ & $-1$ & $1$ & $xy$\\
$\mathrm{E_g}                     $ & $2$ & $0$  & $-2$ & $0$ & $0$  & $2$ & $0$  & $-2$ & $0$ & $0$ & $\{-zx, zy\}$\\
$\mathrm{A}_{1\mathrm {u}}$ & $1$ & $1$  & $1$  & $1$ & $1$  & $-1$& $-1$ & $-1$ & $-1$ & $-1$ & $xyz(x^2-y^2)$\\
$\mathrm{A}_{2\mathrm {u}}$ & $1$ & $1$ & $1$  & $-1$ & $-1$ & $-1$ & $-1$ & $-1$ & $1$ & $1$ & $z$\\
$\mathrm{B}_{1\mathrm {u}}$ & $1$ & $-1$ & $1$ & $1$ & $-1$ & $-1$ & $1$ & $-1$ & $-1$ & $1$ & $xyz$ \\
$\mathrm{B}_{2\mathrm {u}}$ & $1$ & $-1$ & $1$ & $-1$ & $1$ & $-1$ & $1$ & $-1$ & $1$ & $-1$ & $z(x^2-y^2)$\\
$\mathrm{E_u}                     $ & $2$ & $0$ & $-2$ & $0$ & $0$ & $-2$ & $0$ & $2$ & $0$ & $0$ & $\{x,y\}$\\
\hline 
\end{tabular}
\end{center}
\end{table}
\twocolumn

\begin{table}[htbp]
\caption{
\textcolor{black}{
Table of transformation matrices for spin basis and subbands indices.
}
}
\begin{center}
\begin{tabular}{c|c}
\hline
&$(E,H)\otimes (\uparrow,\downarrow)$\\
\hline
$E$ & $I\otimes I$\\
$2C_4$ & $\frac{1}{\sqrt{2}}I\otimes (I + \mathrm{i}\sigma_z)$, $\frac{1}{\sqrt{2}}I\otimes (I - \mathrm{i}\sigma_z)$\\
$C_4^2$ & $\mathrm{i} I\otimes \sigma_z$\\
$2C_2'$ &$\mathrm{i}I\otimes \sigma_x$, $\mathrm{i}I\otimes \sigma_y$\\
$2C_2''$ & $\frac{\mathrm{i}}{\sqrt{2}}I\otimes(\sigma_x + \sigma_y)$, $\frac{\mathrm{i}}{\sqrt{2}}I\otimes(\sigma_x - \sigma_y)$\\
$I$ & $\tau_z\otimes I$\\
$2IC_4=2S_4$ &
       $\frac{1}{\sqrt{2}}\tau_z\otimes (I + \mathrm{i}\sigma_z)$, $\frac{1}{\sqrt{2}}\tau_z\otimes (I - \mathrm{i}\sigma_z)$\\ 
$\sigma_h$ &
       $\mathrm{i} \tau_z\otimes \sigma_z$\\
$2\sigma_v$ &
       $\mathrm{i}\tau_z\otimes \sigma_x$, \textcolor{black}{$\mathrm{i}\tau_z\otimes \sigma_y$}\\
$2\sigma_d$ &
       $\frac{\mathrm{i}}{\sqrt{2}}\tau_z\otimes(\sigma_x + \sigma_y)$, \textcolor{black}{$\frac{\mathrm{i}}{\sqrt{2}}\tau_z\otimes(\sigma_x - \sigma_y)$}\\
\hline
\end{tabular}
\end{center}
\end{table}

\begin{table}[htbp]
\caption{
\textcolor{black}{
Transformation for spatial coordinates.
}
}
\begin{center}
\begin{tabular}{c|c}
\hline
& $(x,y,z)$\\
\hline
$E$
& $(x,y,z)$ \\
$2C_4$
& $(y,-x,z)$, $(-y,x,z)$ \\
$C_4^2$ 
& $(-x,-y,z)$ \\
$2C_2'$ 
& $(x,-y,-z)$, $(-x,y,-z)$\\
$2C_2''$ 
& $(y,x,-z)$, $(-y,-x,-z)$ \\
$I$ 
& $(-x,-y,-z)$ \\
$2IC_4=2S_4$
& $(-y,x,-z)$, $(y,-x,-z)$ \\ 
$\sigma_h$ &
$(x,y,-z)$\\
$2\sigma_v$ &
$(-x,y,z)$, $(x,-y,z)$\\
$2\sigma_d$ &
$(-y,-x,z)$, $(y,x,z)$\\
\hline
\end{tabular}
\end{center}
\end{table}

\subsection{Reference \crd{s}ystems}
\textcolor{black}{When we analyze impurity effects on
2D superconductivities specific to
\crd{those} arising in \crd{the} surface states of 3D TIs,
we need to study
\crd{those} on \crd{the} 2D SC in a topologically trivial reference system.
As a simple reference system, here we introduce a 2D tight-binding model on a square lattice.}

The Hamiltonian of
\textcolor{black}{the} 
2D tight\crd{-}binding model
\textcolor{black}{with an $s$-wave attractive interaction is defined as}
\begin{eqnarray}
\mathcal{H}
&=&\nonumber
t\sum_{\langle i,j\rangle,\sigma}c^\dagger_{i\sigma} c_{j\sigma} 
\textcolor{black}{-g\sum_{i} c^\dagger_{i\uparrow}c_{i\uparrow}c^\dagger_{i\downarrow}c_{i\downarrow}}
\nonumber\\
&+&
u\sum_{i=1}^{N_\mathrm{i}}c^\dagger_{n(i)\sigma}c_{n(i)\sigma}
-
\mu\sum_{i\sigma}c^\dagger_{i\sigma}c_{i\sigma},\label{A4}
\end{eqnarray}
where $t$\textcolor{black}{, $g$, $u$, and} $\mu$ are the hopping between
nearest-neighbor sites\textcolor{black}{,
the amplitude of the on-site attractive interaction,
the impurity potential, and the chemical potential of the system, respectively.}
Here, $n(i)$ is the location of the $i$-th impurity and $N_\mathrm{i}$ is the number of impurities.
\rd{When we consider magnetic scattering
(polarized along \crd{the} $z$\crd{-}axis), we substitute 
\begin{eqnarray}
\rd{u_s}\sum_{i=1}^{N_\mathrm{i}}(c^\dagger_{n(i)\uparrow}c_{n(i)\uparrow}-c^\dagger_{n(i)\downarrow}c_{n(i)\downarrow}),
\end{eqnarray}
for the impurity term in eq.(\ref{A4}),
\begin{eqnarray}
u\sum_{i=1}^{N_\mathrm{i}}c^\dagger_{n(i)\sigma}c_{n(i)\sigma}.
\end{eqnarray}
}

\rd{Similarly to the effective model for Bi$_2$Se$_3$, we introduce
a mean-field decoupling with
real-space order parameters $\Delta_i$ \bl{being} self-consistently} defined by
\begin{eqnarray}
\Delta_{i} = g\langle c_{i\uparrow}c_{i\downarrow}\rangle \label{6.9}.
\end{eqnarray}
\textcolor{black}{\crd{Note} that, when}
$u=0$, the order parameter $\Delta_i$ is homogeneous and
obtained
\crd{using}
\begin{eqnarray}
\Delta_i &\equiv& \Delta_0, \nonumber\\
\Delta_0 
&=& \frac{g}{S}\sum_{\nu}\frac{\Delta_0}{4\sqrt{\rd{E_{\nu}^2}+\Delta_0^2}}. \\
&&
\nonumber
\end{eqnarray}

\subsection{Method \crd{of} \crd{c}ompensating \crd{for c}hange in \crd{d}ensity of \crd{s}tates}
In this subsection, we introduce a
\textcolor{black}{simple}
scheme
\crd{for reducing}
\textcolor{black}{finite\crd{-}size effects in impurity effects on SC.
A dominant finite size effect comes from}
the change in the DOS \textcolor{black}{due to \crd{the} introduced impurity potentials, which is
\underline{assumed} to be negligible in a thermodynamic limit}.
Then, we test the validity of the method by analyzing numerical results on the $s$-wave SC in
\textcolor{black}{the} tight-binding model on the square lattice\textcolor{black}{, introduced in the previous subsection,}
in comparison with the AG theory.

The change in the DOS is \textcolor{black}{assigned as a higher\crd{-}order effect and} 
neglected in the AG theory for continuum models.
\textcolor{black}{However,} in finite-size BdG calculations,
the change in the DOS arising from the impurities quantitatively
affects the order parameter\rd{, in addition to relaxation or pair-breaking processes due to impurities,} 
while this
\rd{\bl{is} assumed to be negligible.} 
Therefore, we have to compensate \crd{for} this change
when one wishes to estimate the genuine reduction
\crd{in} the order parameter purely arising from impurity relaxation processes.
\textcolor{black}{Here we remind the readers that the}
change in the DOS and the effects of the impurity relaxation processes correspond to the diagram in Fig\crd{s}. 1(a) and \crd{1}(b), respectively.

When we neglect such \textcolor{black}{a}
change in the DOS, the pair potential with impurities is calculated in the perturbation regime as
\begin{eqnarray}
\Delta(u,n_{\mathrm i}) &=& \Delta_0 + \delta\Delta(u,n_{\mathrm i}), \label{3.41} \\
\delta\Delta(u,n_{\mathrm i}) &=& -\frac{\pi}{4\tau_s(u,n_{\mathrm i})},\label{3.42}
\end{eqnarray}
according to the AG theory\rd{\cite{cit:JETP12_1243,cit:PR137_A1151}}
where $u$ and $n_{\mathrm{i}}$ is the strength of impurity potentials and the impurity concentration\crd{, respectively}.
In eq.(\ref{3.41}), $\Delta_0$ is the impurity-free pair potential. 
In eq.(\ref{3.42}), $\tau_s(u,n_\mathrm{i})$ is the impurity relaxation time contributing to the reduction
\crd{in} the order parameter.
This relaxation time is determined
\crd{by} the strength of impurity potentials $u$ and the impurity concentration $n_\mathrm{i}$ as well as
the symmetries of the impurity scattering
and
order parameters\cite{cit:RMP78_373,cit:JETP12_1243}.

\textcolor{black}{However,}
in numerical solutions of the BdG equation on
finite-size systems,
\textcolor{black}{the reduction
\crd{in} the order parameter is not fully given by $\delta \Delta (u, n_{\rm i})$ in eq.(\ref{3.41}).}
\textcolor{black}{We need to take into account} the changes in the DOS
as well.
In order to subtract \textcolor{black}{the reduction due to} the changes in the DOS,
we introduce a ``relaxation-ignored" pair potential $\Delta_\mathrm{(DOS)}$,
in which only the change in the DOS by the impurities is taken into account\crd{,} while the impurity relaxation times are neglected.
By using the relaxation-ignored pair potential $\Delta_\mathrm{(DOS)}$, eq.(\ref{3.41}) is replaced by
\begin{eqnarray}
\Delta(u,n_\mathrm{i}) = \Delta_{(\mathrm{DOS})}(u,n_\mathrm{i}) + \delta\Delta(u,n_\mathrm{i}),\label{3.43}
\end{eqnarray}
\rd{when 
\begin{eqnarray}
\frac{u^2n_\mathrm{i}N_0}{\Delta(u,n_\mathrm{i})}\ll 1 \label{A.12}
\end{eqnarray}
holds, where $N_0$ is the DOS at the Fermi energy.
Here, $u^2n_\mathrm{i}N_0$, the numerator
\crd{on} the left\crd{-}hand side
\crd{of} eq.(\ref{A.12}), is the same order as $\delta\Delta(u,n_\mathrm{i})$.}

Then, we introduce an equation by which the relaxation-ignored pair potential is calculated and verify the validity of the method by analyzing $s$-wave SC in a tight-binding model for the square lattice.
\bl{In the} regime of small impurity concentration,
\rd{we define} the relaxation-ignored pair potential $\Delta_\mathrm{(DOS)}$
from
\crd{the} self-consistent equation
\begin{eqnarray}
\Delta_\mathrm{(DOS)}(u,n_\mathrm{i}) = \frac{g}{S}\sum_{\nu}\frac{\Delta_\mathrm{(DOS)}(u,n_\mathrm{i})}{4\sqrt{E_{\nu}(u,n_\mathrm{i})^2+\Delta_\mathrm{(DOS)}(u,n_\mathrm{i})^2}},\label{3.44} \nonumber\\
\end{eqnarray}
where \bl{$E_\nu(u,n_\mathrm{i})$} is the set of eigenvalues when the system has no attractive interaction but has impurities.
By the estimation
\crd{using} eq.(\ref{3.44}), we consider the change in the order parameter due to the shift
\crd{in} the energy spectrum.
The relaxation processes are not contained in the estimate of $\Delta_\mathrm{(DOS)}$, because they appear as
shifts in the imaginary parts of the quasiparticle energies.

The
\crd{estimation} of the relaxation-ignored pair potential by eq.(\ref{3.44}) enables us to compensate
\crd{for} the change in the DOS in the BdG calculations.
When we focus on the relative reduction
\crd{in} the pair potential purely from the impurity relaxation processes,
we should concentrate on \crd{the}
quantity ${\delta\Delta(u,n_\mathrm{i})}/{\Delta_0}$\crd{,} which represents the reduction.
By using eq.(\ref{3.43}), we introduce an equation
\crd{that} associate\crd{s} the results
\crd{of} the BdG calculations with the reduction of the pair potential
\textcolor{black}{purely due to the relaxation processes}. 
The equation is 
\begin{eqnarray}
1+\frac{\delta\Delta(u,n_\mathrm{i})}{\Delta_0}= \frac{\langle\langle\Delta(u,n_\mathrm{i})\rangle_x\rangle_\mathrm{imp}}{\langle\Delta_\mathrm{(DOS)}(u,n_\mathrm{i})\rangle_\mathrm{imp}} \label{3.45},
\end{eqnarray}
which is valid in the
\crd{ranges} of small $n_\mathrm{i}$ and $u$.
Since the pair potentials obtained by the BdG calculations have spatial and impurity-configuration dependence\crd{s}, we take two types of averages\crd{:}
$\langle\cdots\rangle_x$ and $\langle\cdots \rangle_\mathrm{imp}$.
The average $\langle\cdots\rangle_x$ means the spatial average, i.e.,
\begin{eqnarray}
\langle A \rangle_x = \frac{1}{S}\sum_{i=1}^{S}A_i,
\end{eqnarray}
where $S$ is the system size and $A_i$ is a quantity depending on the site $i$.
The other average $\langle\cdots \rangle_\mathrm{imp}$ represents the average over impurity configurations, i.e.,
\begin{eqnarray}
\langle B \rangle_\mathrm{imp} = \frac{1}{N_\mathrm{c}}\sum_{j=1}^{N_\mathrm{c}}B_j ,
\end{eqnarray}
where $N_\mathrm{c}$ is the number of impurity configurations and $B_j$ is a quantity depending on the impurity configuration $j$.
Moreover,
\crd{note} that we have to take $\Delta_\mathrm{DOS}$ after the impurity-configuration average, since it also depends on the impurity configuration.

\textcolor{black}{Then we can compare the results of the BdG with \crd{those obtained using} the AG theory. By using}
the AG theory, the left side of eq.(\ref{3.45}) is estimated as
\begin{eqnarray}
1+\frac{\delta\Delta(u,n_\mathrm{i})}{\Delta_0}= 1-\frac{\pi}{4\tau_s(u,n_\mathrm{i})\Delta_0}.\label{3.46}
\end{eqnarray}
In the cases of the TRS
and magnetic scatterings, the relaxation times $\tau_s$ are obtained
\crd{using}
\begin{eqnarray}
\frac{1}{\tau^\mathrm{(TRS)}_s(u,n_i)} = \mathcal{O}(n_i^2),
\end{eqnarray}
and
\begin{eqnarray}
\frac{1}{\tau^\mathrm{(mag)}_s(u,n_i)} = 2\pi n_i u^2 N_0,
\end{eqnarray}
respectively, where $N_0$ is the DOS at the Fermi energy without impurities, i.e.,
\begin{eqnarray}
N_0 &=& \frac{1}{\pi S}\sum_{\nu}\delta(\epsilon_{0\nu\uparrow})=\frac{1}{2\pi S}\sum_{\nu}\delta(\epsilon_{0\nu}).
\end{eqnarray}
\textcolor{black}{Here\crd{, note}
that, in the finite size system, the \crd{DOS}
is not well-defined. However, we can introduce a reasonable estimation
\crd{of}
the DOS $N_0$
by reconsidering how the \crd{DOS}
$N_0$ appears in the AG theory.
In the AG theory, the
quantity $N'_0$ is replaced by the the DOS in the
thermodynamic limit}
\begin{eqnarray}
N'_0 = \frac{1}{2\pi S}\int^{\bar\omega}_0 \frac{d\omega}{\bar\omega} \frac{\omega}{\omega^2+\epsilon_{0\nu}^2},
\end{eqnarray}
where $\tilde\omega$  is
\crd{the} energy cutoff. 
We adopt $\tilde\omega$ as the bandwidth, i.e., $\tilde\omega=8$\textcolor{black}{,
and estimate the DOS as $N_0 = N'_0$ even in
finite\crd{-}size systems}.

By comparing the two estimate of
$1+\frac{\delta\Delta(u,n_\mathrm{i})}{\Delta_0}$ by eqs. (\ref{3.44}) and (\ref{3.45}),
we
\rd{show}
the validity of our method \crd{of} compensating \crd{for} the change in the
DOS.
Figures A$\cdot$1 and A$\cdot$2 show
the scattering strength and impurity concentration dependence\crd{s} of the pair potential\crd{, respectively}.
By the analyses of
both \bl{dependences}, we find
\crd{an} agreement between the two different approaches\crd{:} the AG theory and
real-space BdG calculation.
Therefore, we conclude that our method of compensating \crd{for} the change in the DOS is valid for analyzing
the impurity concentration dependence of the order parameter.

\begin{figure}[htbp]
\begin{center}
\includegraphics[width=80mm]{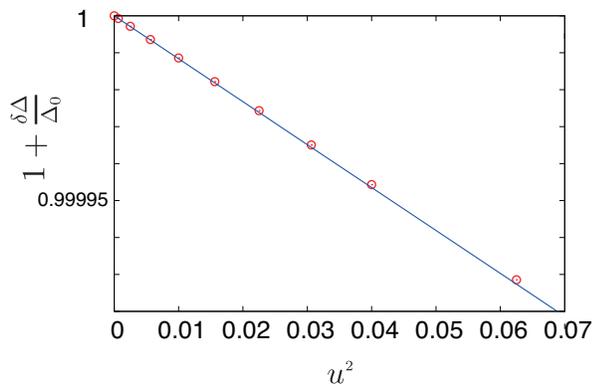}
\caption{(Color online) {Scattering strength} dependence of pair potential for magnetic scattering. 
The {abscissa} represents \crd{the} square of the strength of the impurity potential $u$,
while the {ordinate} represents {$1+\delta\Delta(u,n_\mathrm{i})/\Delta_0$}, which corresponds to
\crd{a} relative change {in} the pair potential.
We take the system size \crd{as} $N_x=N_y=20$, $N_\mathrm{i}=1$, \crd{and} $n_\mathrm{i}=N_\mathrm{i}/(N_xN_y)=0.25 \%$.
The solid line is
\crd{a} theoretical {estimate} by the AG theory. 
The circle symbol corresponds \rd{to the solution of} the BdG equations.}
\label{}
\end{center}
\end{figure}

\begin{figure}[htbp]
\begin{center}
\includegraphics[width=80mm]{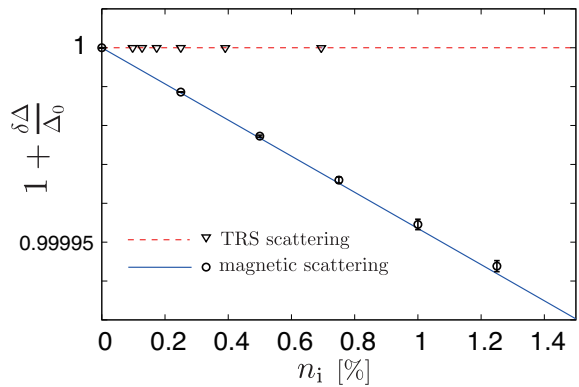}
\caption{(Color online) {Impurity concentration} dependence of the pair potential for TRS and magnetic scatterings. 
The {abscissa} represents the impurity concentration $n_\mathrm{i}$, while the {ordinate} represents {$1+\delta\Delta(u,n_\mathrm{i})/\Delta_0$}.
In the calculation for {the} TRS scattering, we fix the scattering strength \crd{to} $u=0.1$ and the number of impurities \crd{to} $N_\mathrm{i}=1$. 
We alter the impurity concentration by changing the system size $N_xN_y$. 
In the calculation for {the} magnetic scattering, we fix the scattering strength \crd{to} $u=0.1$ and the system size \crd{to} $N_xN_y=400$.
The solid and dashed lines are the theoretical {estimates} for {the} TRS and magnetic scatterings by the AG theory, \bl{respectively}.
The circle and triangle symbols correspond to the results \crd{of} the TRS and magnetic scatterings by the BdG calculations\crd{, respectively}.}
\label{}
\end{center}
\end{figure}

\end{document}